\documentclass[10pt,twosides,journal]{IEEEtran}
\usepackage{algorithm}
\usepackage{algpseudocode}
\usepackage{amsmath}
\usepackage{bm}
\usepackage{multirow}
\usepackage{amsfonts}
\usepackage{cite}
\usepackage[colorlinks,citecolor=red]{hyperref}
\usepackage{graphicx}
\usepackage{epstopdf}

\begin{document}
% ==================================================
\title{DOA and Polarization Estimation for Non-Circular Signals in 3-D Millimeter Wave Polarized Massive MIMO Systems}

\author{Liangtian~Wan,~\IEEEmembership{Member,~IEEE,}
        Kaihui~Liu,~Ying-Chang Liang~\IEEEmembership{Fellow,~IEEE,}
        and~Tong~Zhu% <-this % stops a space

\thanks{L. Wan is with the Key Laboratory for Ubiquitous Network and Service Software of Liaoning Province, School of Software, Dalian University of Technology, Dalian 116620, China (e-mail: wanliangtian@dlut.edu.cn).}% <-this %
\thanks{K. Liu is with National Key Laboratory of Science and Technology on Communications, University of Electronic Science and Technology of China, Chengdu 611731, China (e-mail: kaihuil@std.uestc.edu.cn).}
\thanks{Y.-C. Liang is with National Key Laboratory of Science and Technology on Communications, University of Electronic Science and Technology of China, Chengdu 611731, China, and also with School of Electrical and Information Engineering, University of Sydney, NSW 2006, Australia (e-mail: liangyc@ieee.org).}.
\thanks{T. Zhu is with Tianjin Institute of Computing Technology, 300000, Tianjin, China (e-mail: zhutongheu@gmail.com).}}% <-this % stops a space
%\thanks{Manuscript received April 19, 2005; revised January 11, 2007.}}

%\markboth{Journal of \LaTeX\ Class Files,~Vol.~X, No.~X, January~XXXX}%
%{Shell \MakeLowercase{\textit{et al.}}: Bare Demo of IEEEtran.cls for Journals}

\maketitle

% ==================================================
\begin{abstract}
%boldmath
In this paper, an algorithm of multiple signal classification (MUSIC) is proposed for two-dimensional (2-D) direction-of-arrival (DOA) and polarization estimation of non-circular signal in three-dimensional (3-D) millimeter wave polarized large-scale/massive multiple-input-multiple-output (MIMO) systems. The traditional MUSIC-based algorithms can estimate either the DOA and polarization for circular signal or the DOA  for non-circular signal by using spectrum search. By contrast, in the proposed algorithm only the DOA estimation needs spectrum search, and the polarization estimation has a closed-form expression. First, a novel dimension-reduced MUSIC (DR-MUSIC) is proposed for DOA and polarization estimation of circular signal with low computational complexity. Next, based on the quaternion theory, a novel algorithm named quaternion non-circular MUSIC (QNC-MUSIC) is proposed for parameter estimation of non-circular signal with high estimation accuracy. Then based on the DOA estimation result using QNC-MUSIC, the polarization estimation of non-circular signal is acquired by using the closed-form expression of the polarization estimation in DR-MUSIC. In addition, the computational complexity analysis shows that compared with the conventional DOA and polarization estimation algorithms, our proposed QNC-MUSIC and DR-MUSIC have much lower computational complexity, especially when the source number is large. The stochastic Cram\'{e}r-Rao Bound (CRB) for the estimation of the 2-D DOA and polarization parameters of the non-circular signals is derived as well. Finally, numerical examples are provided to demonstrate that the proposed algorithms can improve the parameter estimation performance when the large-scale/massive MIMO systems are employed.
\end{abstract}

\begin{IEEEkeywords}
Direction-of-arrival (DOA) and polarization estimation, polarized large-scale/massive multiple-input-multiple-output (LS-MIMO/massive MIMO), three-dimensional (3-D) millimeter wave communication,  circular and non-circular signals, quaternion.
\end{IEEEkeywords}

% ==================================================
\section{Introduction}
%\IEEEPARstart

Driven by the proliferation of more sophisticated smart-phone and social media, the wireless mobile traffic will continue to grow at an exponential pace, thus the capacity of cellular data networks needs to increase in orders of magnitude \cite{Akdeniz2014Millimeter}. Millimeter wave communication is a very promising approach for meeting this challenge because of two reasons. First, there is a huge amount of available spectrum ranging from 30 GHz to 300 GHz, which is much more than those used by the existing wireless communication systems. Second, due to the small carrier wavelength of millimeter wave, a large number of antenna elements can be arranged in highly directive steerable arrays \cite{Marzi2016Compressive}. Thus the energy efficiency can be improved dramatically when more energy is concentrated in a particular direction. Furthermore, the highly directive steerable array mentioned above, are also known as massive multiple-input-multiple-output (MIMO), can achieve extremely high spectrum efficiency, thus is the key enabling technology for gigabit-per-second data transmission in millimeter wave communication \cite{Shafin2016DoA}. With the spatial freedom offered by large antenna arrays, abundant mobile terminals are expected to occupy the same set of time and frequency resources with negligible interference \cite{Larsson2014Massive,Cheng2015Subspace}.

Extensive research has been conducted for massive MIMO systems, such as interference mitigation \cite{Yin2013A}, multiuser beamforming \cite{He2014Leakage} and joint spatial division and multiplexing \cite{Adhikary2013Joint}. However, in all the research works mentioned above, knowledge of channel correlations at the base stations (BSs) is required \cite{Cheng2015Subspace}. In order to model channel correlations, the geometric stochastic channel model is widely used \cite{Yin2013A,Adhikary2013Joint,Alkhateeb2014Channel,heath2016overview,Andrews2017Modeling}, wherein direction-of-arrivals (DOAs) of signal paths are crucial model parameters. Thus, accurate DOA estimation for dominant signal paths is a prerequisite for channel correlation acquisitions in millimeter wave communication.

In wireless communication systems, modulated signals based on binary phase shift keying (BPSK) and amplitude modulation (AM) have been widely used. Different from quadrature amplitude modulation (QAM) and quadrature phase shift keying (QPSK) signals, the aforementioned modulated signals are non-circular in the sense that their unconjugated covariance matrices are not equal to zero. This distinctive characteristic can be utilized to improve the parameter estimation performance.

\vspace{-1em}

\subsection{Related Work}
For DOA estimation of non-circular signals, the so-called non-circular multiple signal classification (NC-MUSIC) algorithm was proposed in \cite{Gounon1998Localisation,Abeida2006MUSIC}. The second-order asymptotically minimum variance (AMV) algorithms were proposed in \cite{Delmas2004Asymptotically}, in which a closed-form expression of the lower bound on the asymptotic covariance of estimations  given by arbitrary second-order algorithms was evaluated. However, the computational complexity of these algorithms is tremendous because of the multidimensional nonlinear optimization \cite{Delmas2004Asymptotically}. To reduce the computational complexity, the root-NC-MUSIC algorithm was proposed in \cite{Charge2001A}. Based on a determinant-based method, the DOAs of non-circular and circular signals were simultaneously estimated in \cite{Liu2012A}. However, when two types of signals are too close, the DOA estimation performance may degrade. By exploiting the non-circularity, the DOAs of non-circular and circular signals were separately estimated in \cite{Gao2013Improved}.  By exploiting the stronger orthogonality in the biquaternion domain, the biquaternion cumulant-MUSIC has been proposed for DOA estimation \cite{Gou2013Biquaternion}. Recently, the sparse representation based method has been proposed in \cite{Liu2012Direction} with high estimation accuracy and resolution. However, the computational complexity of this method is much larger than that of the subspace-based methods.

%In [6], the subspace-based estimation algorithms were extended to the case of non-circular signals, and the asymptotic performance of arbitrary subspace-based algorithms has been investigated. However, as pointed out in [7], the method in [6] has a drawback that when non-circular signals (including maximal and common non-circularity rated signals) and circular signals coexist, i.e., in the case of low SNR, limited snapshots and/or large number of sources, more than one spectrum peaking may appear in the neighborhoods of DOAs of common non-circularity rated and circular signals with a certain probability. Another drawback is that when the maximal non-circularity rated signal gets closer to the circular signal, the algorithm cannot resolve these two signals. This is because that when the maximal non-circularity rated signals are estimated, the common non-circularity rated and circular signals are all estimated simultaneously. Theoretical and approximate interpretable closed-form expressions of the array signal-to-noise ratios (ASNR) threshold at which conventional and NC-MUSIC algorithms are able to resolve two closely spaced sources along the Cox and the Sharman and Durrani criteria were given in [8]. An explicit expression of the stochastic Cram\'{e}r-Rao bound (CRB) on DOA estimation accuracy for non-circular sources has been derived in [9]. Recently, the sparse representation based method has been proposed in [10] with high estimation accuracy and resolution. However, the computational complexity of this method is much larger than that of the subspace-based methods.

The one-dimensional (1-D) NC Standard estimation of signal parameters via rotational invariance techniques (ESPRIT) and two-dimensional (2-D) NC Unitary ESPRIT have been proposed in \cite{Zoubir2003Non} and \cite{Haardt2004Enhancements}, respectively, for DOA estimation of non-circular signals. Recently, $R$ multidimensional ESPRIT-type algorithms have been proposed in \cite{Roemer2014Analytical}, and the perturbation analysis of tensor-ESPRIT-type algorithms have been presented as well. Based on this, $R$ multidimensional ESPRIT-type algorithms have been applied to estimate strictly second-order non-circular sources \cite{Steinwandt2014R}, which is regarded as an extension of methods in \cite{Zoubir2003Non} and \cite{Haardt2004Enhancements}. In addition, in \cite{Steinwandt2014R}, the performance of these ESPRIT-type algorithms has been analyzed as well. Recently, two ESPRIT-based algorithms, termed CNC Standard ESPRIT and CNC Unitary ESPRIT, have been devised in \cite{Steinwandt2015ESPRIT} under coexistence of circular and strictly non-circular signals based on NC-ESPRIT methods \cite{Roemer2006Efficient}. They yield closed-form estimation with low computational complexity. However, these methods \cite{Zoubir2003Non,Haardt2004Enhancements,Roemer2014Analytical,Steinwandt2014R,Steinwandt2015ESPRIT} cannot be used for polarization parameter estimations.

For DOA and polarization estimation, the ESPRIT algorithm has been used in the polarization sensitive array. The polarization characteristic of the signal and the relative invariance between the orthogonal dipole and the magnetic output have been exploited for the parameter estimation with uniform linear array (ULA) \cite{Li1991Angle}.  The root-MUSIC algorithm has been proposed based on the diversely polarized characteristic \cite{Wong1996Diversely}. In the case when the array manifold is partly known, the fourth-order statistics-based method has been presented for joint parameter estimation \cite{Gonen1999Applications}. Then 2$q$th-order, $q \ge 2$, MUSIC methods have been applied to arrays having diversely polarized antennas for diversely polarized sources \cite{Chevalier2007Higher}. The parallel factor (PARAFAC) analysis (low order tensor) has been used for estimating DOA and polarization parameters, and the conventional complex matrix model is replaced by the low order tensor model \cite{Liu2001Cramer}. The orthogonality among propagation direction of electromagnetic wave, electric field and magnetic field is reflected profitably \cite{Guo2011A}. Based on the effective aperture distribution function, an extension of root-MUSIC algorithm was proposed for DOA and polarization estimation with arbitrary array configurations \cite{Costa2012DoA}. The sparse representation based method has been proposed in \cite{Tian2015Sparse} for ULA by solving a weighted group lasso problem in second-order statistics domain. However, the computational complexity of this method is much larger than that of the subspace-based methods. The quaternion-MUSIC algorithm has been proposed in \cite{Miron2006Quaternion}, and a comparison between long vector orthogonality and quaternion vector orthogonality is also performed.  The biquaternion matrix diagonalization has been used for DOA estimation based on vector-antennas \cite{Bihan2013MUSIC,Gong2011Direction}. However, these algorithms assume that the signal is circular, the estimation performance of non-circular signal cannot be improved any further. To the best of our knowledge, no contributions have dealt yet with DOA and polarization for non-circular signals.

\vspace{-1em}
\subsection{Movitation}
Recently, there has been a gradual demand for the use of polarized antenna systems, especially for 5G mobile communication systems \cite{Poon2011Degree,Su2016Channel}. This is because of the fact that, for the design of space-limited wireless devices, the antenna polarization is a crucial resource to be exploited. The degree-of-freedom and multiplexing could be increased by exploiting the antenna polarization. In addition, a massive MIMO system equipped with electromagnetic vetor sensors (EMVSs) could generally form a uniform rectangular array (URA).  It should be noted that this URA could estimate not only the DOA of the incident signal, but also its polarization. The BS could use polarization parameters to distinguish different mobile terminals, since those parameters should contain unique identification of mobile terminals. In a secure millimeter wave communication, the polarization parameters can be used for encrypting the classified information, only the polarized massive MIMO systems could decode this encryption information. There should be other applications that the polarization  parameters could be used for. Thus, the polarization parameters' estimation using polarized massive MIMO systems is a meaningful research field in millimeter wave communication as well.

In this paper, we adopt the polarized massive MIMO systems to estimate the 2-D DOA and polarization of multiple no-circular sources, since accurate DOA and polarization estimations are particularly critical for channel correlation acquisitions, as well as for the mobile terminal identification and the information security mentioned above in millimeter wave communication.

\vspace{-1em}
\subsection{Contribution}
In this paper, a MUSIC-based algorithm is proposed for 2-D DOA and polarization estimation of multiple no-circular sources in polarized massive MIMO systems employing very large URAs. The circular signal model containing DOA and polarization parameter are constructed for polarized massive MIMO systems, and the non-circular signal model is constructed based on quaternion theory. The partial derivative of the spectrum function is utilized to reduce the dimension of parameter search. The DOA parameter is estimated at first, and the polarization parameter is estimated based on the results of the DOA parameter. To be more specific, the main contributions of this paper are listed as follows.

1) For circular signals, a dimension-reduced MUSIC (DR-MUSIC) algorithm is proposed for DOA and polarization estimation. Compared with classical long-vector MUSIC (LV-MUSIC) and quaternion dimension-reduced MUSIC (QDR-MUSIC) algorithm \cite{Li2011The}, the computational complexity of DR-MUSIC algorithm is further reduced, since the polarization estimation of DR-MUSIC has a closed-form expression.

2) For non-circular signals, an improved DOA estimation algorithm is proposed based on the URA equipped with EMVSs. Compared with the QDR-MUSIC algorithm, the estimation accuracy is further improved. This is because a novel received data model is constructed based on quaternion theory, and the unconjugated covariance matrix of non-circular has been used in the proposed quaternion non-circular MUSIC (QNC-MUSIC) algorithm to improve the DOA estimation accuracy.

3) By combining the QNC-MUSIC and DR-MUSIC algorithms, the polarization estimation can be achieved for non-circular signals. Based on the result for the DOA estimation using QNC-MUSIC algorithm, the polarization estimation of non-circular signal can be acquired by using the closed-form expression of the polarization estimation of DR-MUSIC algorithm.

4) The computational complexity of the LV-MUSIC, DR-MUSIC, QDR-MUSIC and QNC-MUSIC are analyzed. Compared with LV-MUSIC and QDR-MUSIC, the computational complexity of DR-MUSIC and QNC-MUSIC is much lower. This advantage is particularly attractive in the massive MIMO systems, since the potentially prohibitive computational complexity is one of the major challenges faced by massive MIMO systems.

5) The stochastic Cram\'{e}r-Rao Bound (CRB) for the estimation of the 2-D DOA and polarization parameters of the non-circular signals is derived, whereas the known CRB is only valid for the estimation of the 2-D DOA and polarization parameters of the circular signals.
\vspace{-1em}
\subsection{Organization of the Paper}

This paper is organized as follows. The problem formulations are given in Section II. The basic concept and property of quaternion are given in Section III. The proposed DR-MUSIC algorithm for circular signal is presented in Section IV.  The proposed QNC-MUSIC algorithm for non-circular signal is presented in Section V. The computational complexity analysis is given in Section VI. The stochastic CRB is derived in Section VII. The simulation results are shown and analyzed in Section VIII. The conclusions are drawn in Section IX.

\vspace{-1em}
\subsection{Notation}
In this paper, the operator ${\left(  \cdot  \right)^\dag }$, ${\left(  \cdot  \right)^ * }$, ${\left(  \cdot  \right)^T}$, ${\left(  \cdot  \right)^H}$ and $\mathrm{E}\left\{  \cdot  \right\}$ are  complex matrix pseudo-inverse, conjugate, transpose, conjugate transpose and expectation, respectively; the operator ${\left(  \cdot  \right)^\# }$, ${\left(  \cdot  \right)^\diamond }$,  ${\left(  \cdot  \right)^\ddag }$ and $\mathbb{E}$ are conjugate, transpose, conjugate transpose and expectation for quaternion matrix, respectively. The boldface uppercase letters and boldface lowercase letters denote matrices and column vectors, respectively. The symbol diag$\left\{ {{z_1},{z_2}} \right\}$ stands for a diagonal matrix whose diagonal entries are ${z_1}$ and ${z_2}$, respectively.  ${\bm{I}_M}$ and $\bm{J}_{M}$ stand for the $ M \times M $ identity matrix and the $ M \times M $ matrix of ones, respectively. $\left|  \cdot  \right|$ and $\left\| \cdot \right\|$ stand for the module operator and the absolute value operator, respectively. $\left\| \cdot \right\|_F$ denotes Frobenius norm. $\mathrm{arg}(\cdot)$ is the phase operator of complex numbers, in radian. Symbols $\odot$ and $\otimes$ stand for the Hadamard matrix product and the Kronecker product, respectively. $\perp$ denotes the ortho-complement of a projector matrix.

\vspace{-1em}
\section{Problem Formulation}
\begin{figure}
\centering
\includegraphics[width=3in,height=1.9in]{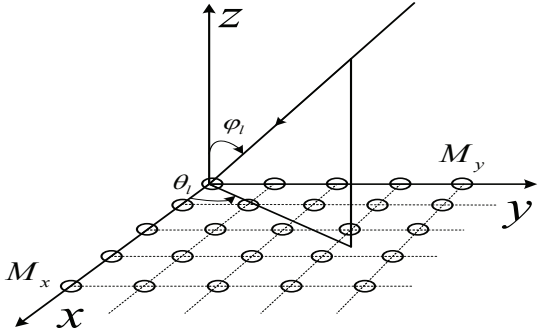}
\vspace{-1em}
\caption{Array geometry of the URA considered. The direction of the incident path is projected onto the array plane. The azimuth angle, ${\theta _l}$, is defined as the angle from the x-axis to the projected line, and the elevation DOA, ${\varphi _l}$, is defined as the angle from the z-axis to the incident path. The ranges of the two parameters
are $0\leq \theta _l< \pi$ and $0\leq \varphi _l< \pi/2$.}\label{Fig1}
\vspace{-2em}
\end{figure}

As shown in Fig. \ref{Fig1}, we consider a 3-D millimeter wave polarized  massive MIMO system with EMVSs arranged in a URA form at the BS. There are totally $M=M_xM_y$ EMVSs, where $M_x$ and $M_y$ are the numbers of antennas in the $x$-direction and the $y$-direction, respectively. Obviously, the URA would be degenerated to the conventional ULA when $M_x$ or $M_y$ are equal to $1$.

For the mobile terminals, each is equipped with one EMVS. The uplink signals of the $L$ mobile terminals are non-circular signals such as BPSK modulated signals going through $L$ channels, and each has a corresponding azimuth angle ${\theta _l}$ and elevation angle ${\varphi _l}$ for the $l$th mobile terminal, which satisfy $0\leq \theta _l< \pi$ and $0\leq \varphi _l< \pi/2$. The $L$ channels are uncorrelated with each other. It should be noted that the ranges of the DOAs are the localization ranges of the array, which means that sources out of these ranges cannot be localized by the array. The steering vector $\bm{a}(\theta _l,\varphi _l)\in\mathbb{C}^{M\times1}$ is the response of the array corresponding to the azimuth and elevation DOAs of ${\theta _l}$ and ${\varphi _l}$. With respect to the EMVS at the origin of the axes, the $m$th element of $\bm{a}(\theta _l,\varphi _l)$ is defined as
\begin{equation}\label{eq1}
\begin{aligned}
\left [\bm{a}(\theta _l,\varphi _l)  \right ]_m=&\exp(iu\sin\varphi_l[(m_\mathrm{x}-1)\cos\theta _l\\
&+(m_\mathrm{y}-1)\sin\theta_l] ),\\
\end{aligned}
\end{equation}
where $m=(m_\mathrm{y}-1)M_\mathrm{x}+m_\mathrm{x}, m_\mathrm{x}=1,2,\dots ,M_\mathrm{x},m_\mathrm{y}=1,2,\dots ,M_\mathrm{y}$, $u=2{\pi}d/\lambda$, $d$ is the distance between two adjacent EMVSs, $\lambda$ is the wavelength. It can be seen that $\left [\bm{a}(\theta _l,\varphi _l)  \right ]_m$ corresponds to the response of the $(m_\mathrm{x},m_\mathrm{y})$th EMVS in the coordinate system shown in Fig. \ref{Fig1}.

An EMVS equipped with two dipole antennas offers a good trade-off between performance and overall system development cost for the polarized massive MIMO systems equipped with a large number of EMVSs, thus we consider the case that an EMVS equipped with two dipole antennas, which are arranged in the $x$-direction and the $y$-direction, respectively, measures the horizontal and vertical components of the electronic field.  For the $l$th channel, the components of the electric field received on an EMVS can be defined as \cite{Li1991Angle}
\begin{equation}\label{components}
\begin{aligned}
\xi_l(\theta_l ,\varphi_l,\gamma_l,\eta_l )=&\begin{bmatrix}
\xi _{1l}(\theta_l ,\varphi_l,\gamma_l,\eta_l )\\
\xi _{2l}(\theta_l ,\varphi_l,\gamma_l,\eta_l )
\end{bmatrix} \\
=&\begin{bmatrix}
\cos\theta_l\cos \varphi_l  & -\sin\theta_l\\
\sin\theta_l\cos \varphi_l & \cos\theta_l
\end{bmatrix}\begin{bmatrix}
\sin\gamma_l \exp(i\eta_l )\\
\cos \gamma_l
\end{bmatrix},
 \end{aligned}
\end{equation}
$l=1,2,\dots,L$, where $\xi _{1l}(\theta_l ,\varphi_l,\gamma_l,\eta_l )$ and $\xi _{2l}(\theta_l ,\varphi_l,\gamma_l,\eta_l )$ stand for the horizontal and vertical components of the electronic field received by an EMVS in the $x$-direction and the $y$-direction, respectively, while the $0\leq \gamma _l<\pi/2$ and $0\leq \eta _l<2\pi$ are the ranges of the polarization angle and phase difference, respectively. Thus the time-domain signals received by the $m$th EMVS equipped with two dipole antennas can be expressed as
\vspace{-1em}
\begin{equation}
\begin{aligned}
{x_{1m}}(t) = \sum\limits_{l = 1}^L {\left [\bm{a}(\theta _l,\varphi _l)  \right ]_m \xi _{1l}(\theta_l ,\varphi_l,\gamma_l,\eta_l ){s_l}(t)} + {n_{1m}}(t),\\
{x_{2m}}(t) = \sum\limits_{l = 1}^L {\left [\bm{a}(\theta _l,\varphi _l)  \right ]_m \xi _{2l}(\theta_l ,\varphi_l,\gamma_l,\eta_l ){s_l}(t)} + {n_{2m}}(t),
 \end{aligned}
\end{equation}
$m = 1,2, \ldots ,M$, where ${{s_l}(t)}$ is the complex envelope of the received signal, ${n_{1m}}(t)$ and ${n_{2m}}(t)$ are the additive white Gaussian noise (AWGN) of the $m$th EMVS consisting of two dipole antennas. It is to be noted here that we will replace $\xi _{1l}(\theta ,\varphi,\gamma,\eta )$, $\xi _{2l}(\theta ,\varphi,\gamma,\eta )$ and $\left [\bm{a}(\theta _l,\varphi _l)  \right ]_m$ with $\xi _{1l}$, $\xi _{2l}$ and $\left [\bm{a}_l  \right ]_m$, respectively, in the following for notational convenience.  Let ${\bm{x}_1}\left( t \right) = {\left[ {{x_{11}},{x_{12}}, \ldots ,{x_{1M}}} \right]^T}$, ${\bm{x}_2}\left( t \right) = {\left[ {{x_{21}},{x_{22}}, \ldots ,{x_{2M}}} \right]^T}$, then the data model that the array received can be expressed as
\begin{equation}\label{outpa}
\begin{aligned}
\overline{\bm{x}}\left( t \right) = \begin{bmatrix}
{{\bm{x}_1}\left( t \right)}\\
{{\bm{x}_2}\left( t \right)}
\end{bmatrix}& = \begin{bmatrix}
{\bm{A}{\bm{V}_1}}\\
{\bm{A}{\bm{V}_2}}
\end{bmatrix}\bm{S}\left( t \right) +\begin{bmatrix}
{{\bm{n}_1}\left( t \right)}\\
{{\bm{n}_2}\left( t \right)}
\end{bmatrix}\\
&=\overline{\bm{A}}\bm{S}(t)+\bm{N}(t),
\end{aligned}
\end{equation}
where $\bm{A} = \left[ {{\bm{a}_1},{\bm{a}_2}, \ldots ,{\bm{a}_L}} \right] \in {\mathbb{C}^{M \times L}}$ is the array manifold matrix,  $\bm{S}\left( t \right) = {\left[ {{s_1}\left( t \right),{s_2}\left( t \right), \ldots ,{s_L}\left( t \right)} \right]^T} \in {\mathbb{C}^{L \times 1}}$ is the signal vector.  ${\bm{V}_k}\left( t \right) = {\rm{diag}}\left\{ {{\xi_{k1}},{\xi_{k2}}, \ldots ,{\xi_{kL}}} \right\} \in {\mathbb{C}^{L \times L}}\left( {k = 1,2} \right)$ is a diagonal matrix constructed by the components of the electric field. ${\bm{n}_k}\left( t \right) = {\left[ {{n_{k1}(t)},{n_{k2}(t)}, \ldots ,{n_{kM}(t)}} \right]^T} \in {\mathbb{C}^{M \times 1}}\left( {k = 1,2} \right)$ are the noise vectors composed of temporally and spatially independent and identically distributed (i.i.d.) circularly symmetric zero-mean Gaussian random variables, whose covariance matrix is $\mathrm{E}\left \{ {\bm{n}_k}\left( t \right){\bm{n}^H_k}\left( t \right) \right \}=\sigma _n^2\bm{I}_M$. It should be noted that the unconjugated covariance matrix of ${\bm{n}_k}\left( t \right)$ is $\mathrm{E}\left \{ {\bm{n}_k}\left( t \right){\bm{n}^T_k}\left( t \right) \right \}=\bm{0}$.

Then the covariance matrix ${\bm{R}_{\bm{x}}}$ of $\overline{\bm{x}}\left( t \right)$ in (\ref{outpa}) can be written as
\begin{equation}\label{DLCOV}
\begin{aligned}
 {\bm{R}_{\bm{x}}}&= \mathrm{E} \left[ \overline{\bm{x}}(t)\overline{\bm{x}}(t)^H \right] \\
 & = \begin{bmatrix}
{{{\bm{A}}}{{\bm{V}}_1}}\\
{{{\bm{A}}}{{\bm{V}}_2}}
\end{bmatrix} \bm{R}_{\bm{s}}(t) \begin{bmatrix}
{{{\bm{A}}}{{\bm{V}}_1}}\\
{{{\bm{A}}}{{\bm{V}}_2}}
\end{bmatrix}^H  + \sigma _n ^2{\bm{I}_{2M}}, \\
 \end{aligned}
\end{equation}
where $\bm{R}_{\bm{s}}(t)=\mathrm{E}[\bm{S}(t)\bm{S}^H(t)]$ is the complex covariance matrix of signal $\bm{S}(t)$.
Collecting  $N$ snapshots ${t_1},{t_2}, \ldots ,{t_N}$, the received data can be written in matrix form as
\begin{equation}
\overline{\bm{X}} = \left[ {\overline{\bm{x}}\left( {{t_1}} \right),\overline{\bm{x}}\left( {{t_2}} \right), \ldots ,\overline{\bm{x}}\left( {{t_N}} \right)} \right].
\end{equation}

The DOA and polarization estimation problem is stated as follows. Given the received data $\overline{\bm{X}}$, we need to estimate the DOA and polarization parameters $\left( {{\theta _l},{\varphi _l},{\eta _l},{\gamma _l}} \right)$, $l = 1,2, \ldots ,L$.

\vspace{-1em}
\section{Basic Concept and Property of Quaternion}
Hamiltion's quaternions $\mathbb{H}$ is a nontrivial generalization of complex numbers $\mathbb{C}$. In general, quaternions are a four dimensional hypercomplex numbers system, and they are an extension of complex numbers to four-dimensional (4-D) space. Basics about quaternions and their properties can be found in \cite{Bihan2004Singular} and \cite{Ward1997Quaternions}. Several basic definitions and properties which would be used in this paper are introduced as follows.

\textbf{Definition 1} \cite{Miron2006Quaternion}: A quaternion  is described by four components (one real and three imaginaries). It can be expressed in its Cartesian form as
\begin{equation}
\begin{aligned}
\alpha  = &{\alpha _0} + {\alpha _1}j_1 + {\alpha _2}j_2 + {\alpha _3}j_3\\
=&({\alpha _0}+{\alpha _1}j_1)+({\alpha _2}+{\alpha _3}j_1)j_2\\
=&c_1+c_2j_2, {\alpha _0},{\alpha _1},{\alpha _2},{\alpha _3} \in \mathbb{R}.
\end{aligned}
\end{equation}
where $j_1j_2=j_3, j_2j_1=-j_3$, $c_1,c_2 \in\mathbb{C}$ are complex numbers. The modulus of a quaternion $\alpha$ can be calculated as $|\alpha|=\sqrt{| c_1|^2+| c_2|^2}$.

The set of quaternions forms a noncommutative normed division algebra, which means that given two quaternions $\alpha$ and $\kappa$, we have $\alpha\kappa\neq\kappa\alpha$.

\textbf{Definition 2} \cite{Miron2006Quaternion}: In this paper, the conjugate of a quaternion $\alpha$, noted $\alpha^\#$, is given by $\alpha^
\#  = {\alpha _0} - {\alpha _1}j_1 - {\alpha _2}j_2 - {\alpha _3}j_3$.

\textbf{Definition 3} \cite{Bihan2004Singular}: The modulus of a quaternion $\alpha$  is $|\alpha|$, which can be expressed as $|\alpha|=\sqrt{\alpha\alpha^\#}=\sqrt{\alpha^2 _0+\alpha^2 _1+\alpha^2 _2+\alpha^2 _3}$.

\textbf{Property 1} \cite{Bihan2004Singular}: For two quaternions $\alpha, \kappa \in \mathbb{H}$ and a complex number $c \in\mathbb{C}$ with its imaginary part $j_1$, noted $c=a+bj_1\in \mathbb{C}$, we have   $(\alpha\kappa )^\#=\kappa^\#\alpha^\#, cj_2=j_2c^*$. Thus the conjugate of $\alpha$ can be written as $\alpha^\#=(c_1+c_2j_2)^\#=c^*_1-j_2c^*_2=c^*_1-c_2j_2$.

\textbf{Remark 1}: It should be noted that the imaginary part $i$ used in section II is essentially identical with the imaginary part $j_1,j_2,j_3$ used in section III. In this paper, $i$ is equivalent to $j_1$.

\textbf{Definition 4}  \cite{Miron2006Quaternion}: Define the $(p,q)$th entry of a $P\times Q$ matrix $\bm{B}$ as a quaternion $[\bm{B}]_{p,q}\in\bm{H}$, then $\bm{B}\in\bm{H}^{P\times Q}$ is called the quaternion matrix.

\textbf{Property 2} \cite{Ward1997Quaternions}: For two quaternion square matrices $\bm{B}, \bm{C} \in \mathbb{H}^{P\times P}$, $\alpha \in \mathbb{H}$ and a complex matrix $\bm{D}\in\mathbb{C}^{P\times Q}$ with its imaginary part $j_1$, we have $(\alpha\bm{C})^\ddag=\bm{C}^\ddag\alpha^\#, (\bm{B}\bm{C})^\ddag=\bm{C}^\ddag\bm{B}^\ddag, \bm{D}j_2=j_2\bm{D}^*$.

It should be noted that, in general, for the transpose of two quaternion matrices $\bm{B}, \bm{C} \in \mathbb{H}^{P\times P}$, we have $(\bm{B}\bm{C})^\diamond\neq\bm{C}^\diamond\bm{B}^\diamond$.

\textbf{Definition 5} \cite{Bihan2004Singular}: Given a quaternion matrix  $\bm{B}\in \mathbb{H}^{P\times Q}$, The Cayley-Dickson notation can be written as $\bm{B} = {\bm{B}_1} + j_2{\bm{B}_2}\left( {{\bm{B}_1},{\bm{B}_2} \in {\mathbb{C}^{P \times Q}}} \right)$. Then one can define the complex adjoint matrix, denoted by $\bm{B}^\sigma \in {\mathbb{C}^{2P \times 2Q}}$, corresponding to the quaternion matrix, as follows
\begin{equation}\label{comrep}
{\bm{B}^\sigma } = \left( {\begin{array}{*{20}{c}}
   {{\bm{B}_1}} & {{\bm{B}_2^*}}  \\
   { - \bm{B}_2 } & {\bm{B}_1^ * }  \\
\end{array}} \right) .
\end{equation}

\textbf{Definition 6} \cite{Zhang1997Quaternions}: Given a quaternion square matrix $\bm{B}\in \mathbb{H}^{P\times P}$, there exists a quaternion $\lambda \in \mathbb{H}$ and a non-zero vector $\bm{b} \in \mathbb{H}^{P\times 1}$, which satisfy $\bm{Bb}=\lambda\bm{b}$ (Left) or $\bm{Bb}=\bm{b}\lambda$ (Right), $\lambda$ is the left or right eigenvalue of the matrix $\bm{B}$, and $\bm{b}$ is the eigenvector of the matrix $\bm{B}$ corresponding to $\lambda$.

%Due to the noncommutative multiplication rule of quaternions, the right eigenvalue does not equal to the left eigenvalues. In this paper. we only use the right eigenvalue and we adopt
%the name 'eigenvalue' to represent the right eigenvalue.

\textbf{Theorem 1} \cite{Zhang1997Quaternions}: Given a quaternion square matrix $\bm{B}\in \mathbb{H}^{P\times P}$ satisfies $\bm{B}=\bm{B}^\ddag$, then $\bm{B}$ is called self-conjugated matrix.  The self-conjugated matrix $\bm{B}$ has equivalent right and left eigenvalues. Moreover, they are both real numbers, and they are the eigenvalues of the complex representation matrix ${\bm{B}^\sigma }$ as well.

\begin{IEEEproof}
This proof can be found in \cite{Zhang1997Quaternions}.
\end{IEEEproof}

For a quaternion self-conjugated matrix $\bm{B} = {\bm{B}_1} + j_2{\bm{B}_2} \in \mathbb{H}^{P\times P}$ with $\bm{B}=\bm{B}^\ddag$, we have $\bm{B}_1=\bm{B}^H_1$ and $\bm{B}_2=-\bm{B}^T_2$. According to (\ref{comrep}), it can be known that $\bm{B}^\sigma$ is a ${2P\times 2P}$ complex Hermite matrix. Based on the eigenvalue decomposition (EVD) of the complex self-conjugated matrix $\bm{B}^\sigma$, the EVD of the corresponding quaternion matrix $\bm{B}$ can be acquired. The EVD of the complex adjoint matrix ${\bm{B}^\sigma }$ can be expressed as
\begin{equation} \label{evd}
{\bm{B}^\sigma }=\begin{bmatrix}
\hat{\bm{U}}_1 & \hat{\bm{U}}_2^*\\
-\hat{\bm{U}}_2 & \hat{\bm{U}}^*_1
\end{bmatrix}\begin{bmatrix}
\hat{\bm{\Lambda}}  & \bm{0}\\
\bm{0} & \hat{\bm{\Lambda}}
\end{bmatrix}\begin{bmatrix}
\hat{\bm{U}}_1 &\hat{ \bm{U}}_2^*\\
-\hat{\bm{U}}_2 & \hat{\bm{U}}^*_1
\end{bmatrix}^H,
\end{equation}
where $\hat{\bm{\Lambda}}=\mathrm{diag}\{\lambda_1,\lambda_2,\cdots,\lambda_P\}$, $\lambda_p, p=1,2,\cdots,P$ are the eigenvalues of quaternion matrix $\bm{B}$. Then the EVD of  a quaternion self-conjugated matrix $\bm{B} = {\bm{B}_1} + j_2{\bm{B}_2} \in \mathbb{H}^{P\times P}$ can be expressed as
\begin{equation}\label{EVD1}
\bm{B}=\hat{\bm{U}}\hat{\bm{\Lambda}}\hat{\bm{U}}^\ddag=(\hat{\bm{U}}_1+j_2\hat{\bm{U}}_2)\hat{\bm{\Lambda}}(\hat{\bm{U}}_1+j_2\hat{\bm{U}}_2)^\ddag.
\end{equation}

\vspace{-1em}
\section{DOA and Polarization Estimation for Circular Signals}
In this section, first, a quaternion dimension-reduced MUSIC (QDR-MUSIC) algorithm proposed in \cite{Li2011The} is introduced; then a dimension-reduced MUSIC (DR-MUSIC) algorithm  is proposed based on the partial derivative of the spectrum function \cite{IEEEhowto:31}.  Compared with QDR-MUSIC, the computational complexity of our proposed DR-MUSIC is further reduced.
%by us in 2014 is introduced \cite{IEEEhowto:31}
\vspace{-1em}
\subsection{Quaternion Dimension-Reduced MUSIC Algorithm}
Based on the quaternion theory, the signal received by an EMVS equipped with two dipole antennas can be combined. The received time-domain quaternion ${x_m}(t) \in \mathbb{H}$ of the $m$th EMVS can be expressed as
\begin{equation}\label{rdata}
\begin{aligned}
 {x_m}(t)  = & {x_{1m}}(t) + j_2{x_{2m}}(t) \\
= & \sum\limits_{l = 1}^L {\left [\bm{a}_l  \right ]_m \xi _{1l}{s_l}(t)}  + {n_{1m}}(t)\\
&+ j_2 \left( {\sum\limits_{l = 1}^L {\left [\bm{a}_l  \right ]_m \xi _{2l}{s_l}(t)}  + {n_{2m}}(t)} \right) \\
 \buildrel \Delta \over = & \sum\limits_{l = 1}^L { \left [\bm{a}_l  \right ]_m{\xi_l}{s_l}(t)}  + {n_m}(t), \\
\end{aligned}
\end{equation}
where  ${\xi_l=\xi _{1l} + j_2\xi _{2l}}\in \mathbb{H}$ is a quaternion, and the quaternion ${n_m}(t)={n_{1m}}(t) + j_2{n_{2m}}(t) \in \mathbb{H} $ is the additive noise of the $m$th EMVS. It follows that
\begin{equation}\label{rv}
\begin{aligned}
{\bm{x}}(t)&={\bm{x}_1}(t)+j_2{\bm{x}_2}(t)\\
&= \sum\limits_{l = 1}^L {{{\bm{a}}_l}{\xi_l}{s_l}(t)}  + {\bm{n}}(t) \buildrel \Delta \over = {{\bm{A}}}{\bm{\xi}}{\bm{s}}(t) + {\bm{n}}(t),
\end{aligned}
\end{equation}
where ${\bm{\xi}} = \mathrm{diag}\{ {{\xi_1},{\xi_2}, \ldots ,{\xi_L}} \} \in \mathbb{H}^{L\times L}$ is a quaternion matrix.  When the self uncorrelated noise is independent of the signals, the covariance matrix of ${\bm{x}}(t)$ that the array received can be expressed as
\begin{equation}\label{aoup}
\begin{aligned}
{\bm{R}} =&\mathbb{ E} \left[ {{\bm{x}}(t)  {{\bm{x}}^\ddag }(t)  } \right] = \bm{A}{\bm{\xi}}{{\bm{R}} }_{\bm{s}}({\bm{A}}{\bm{\xi}}) ^\ddag  + \mathbb{E} \left[ {{\bm{n}}(t){{\bm{n}}^\ddag }(t)  } \right]\\
=&\bm{A}{\bm{\xi}}{{\bm{R}} }_{\bm{s}}({\bm{A}}{\bm{\xi}}) ^\ddag+2\sigma _n ^2{\bm{I}_M},
\end{aligned}
\end{equation}
where ${{\bm{R}}_{\bm{s}} }$ is signal covariance matrix. $\mathbb{E}\left[ {{\bm{n}}(t){{\bm{n}}^\ddag }(t) } \right] =\mathbb{E} [(\bm{n}_1(t) + j_2 \bm{n}_2(t) )( \bm{n}^H_1 (t) - \bm{n}^H_2 (t)j_2) ] = 2\sigma _n ^2{\bm{I}_M}$.

In practical situations, the theoretical array covariance matrices given in (\ref{aoup}) is unavailable and it can be estimated by
\begin{equation}\label{PRACOV}
\begin{aligned}
\hat{\bm{R}}&=\frac{1}{N}\{\bm{X}\bm{X}^\ddag\}=\frac{1}{N}({\bm{X}_1}+j_2{\bm{X}_2})({\bm{X}_1}+j_2{\bm{X}_2})^\ddag\\
&=\frac{1}{N}[\bm{X}_1\bm{X}_1^H+\bm{X}_2^*\bm{X}_2^T+j_2(\bm{X}_2\bm{X}_1^H-\bm{X}_1^*\bm{X}_2^T)]\\
&\buildrel \Delta \over=\bm{R}_1+j_2\bm{R}_2,
\end{aligned}
\end{equation}
where $\bm{X}\in \mathbb{H}^{M\times N}$, $\bm{X}_1\in \mathbb{C}^{M\times N}$ and $\bm{X}_2\in \mathbb{C}^{M\times N}$ are the snapshot data matrices constructed by $\bm{x}(t)$, $\bm{x}_1(t)$ and $\bm{x}_2(t)$, respectively, $t=t_1,\dots,t_N$.

Given the quaternion self-conjugated covariance matrix  $\bm{R}\in \mathbb{H}^{M\times M}$, according to the quaternion form defined in (\ref{rdata}) and the corresponding EVD of its complex adjoint matrix defined in (\ref{evd}) , the EVD of complex Hermite adjoint matrix ${\bm{R}^\sigma }$ is given by
\begin{equation}\label{cam}
 {{\bm{R}}^\sigma }  =\begin{bmatrix}
{{{\bm{R}}_1}}& {{{\bm{R}}_2^*}}\\
{ - {{\bm{R}}_2}} & {{{\bm{R}}_1^*}}
\end{bmatrix} =\begin{bmatrix}
{{{\bm{U}}_1}}& {{{\bm{U}}_2^*}}\\
{ - {{\bm{U}}_2}} & {{{\bm{U}}_1^*}}
\end{bmatrix}\begin{bmatrix}
{\bm{\Lambda }}& \bm{0}\\
\bm{0} & {\bm{\Lambda }}
\end{bmatrix}{\begin{bmatrix}
{{{\bm{U}}_1}}& {{{\bm{U}}_2^*}}\\
{ - {{\bm{U}}_2}} & {{{\bm{U}}_1^*}}
\end{bmatrix}}^H.  \\
\end{equation}

The EVD of quaternion covariance matrix $\bm{R}$ can be expressed as
\begin{equation}\label{covevd}
 {\bm{R}}={\bm{U\Lambda }}{{\bm{U}}^\ddag } = {{\bm{U}}_S }{{\bm{\Lambda }}_S }{\bm{U}}_S ^\ddag  + 2\sigma _n ^2{{\bm{U}}_N }{\bm{U}}_N ^\ddag,  \\
\end{equation}
where ${{\bm{U}}_{N}}={{{\bm{U}}_{N 1}} + j_2 {{\bm{U}}_{N 2}}}\in \mathbb{H}^{M\times (M-L)}$ and ${{\bm{U}}_{N 1}},{{\bm{U}}_{N 2}\in \mathbb{C}^{M\times (M-L)}}$.

Similar to the property of complex matrix MUSIC algorithm \cite{Schmidt1986Multiple}, the quaternion matrix ${{\bm{U}}_S }$ and ${{\bm{U}}_N }$ satisfy the orthogonality relationship, i.e., the steering vector belonging to $\bm{A}{\bm{\xi}}$ is orthogonal to ${{\bm{U}}_N }$.

The spectrum function of quaternion dimension-reduced MUSIC (QDR-MUSIC) is constructed as
\begin{equation}\label{sfQMUSIC}
\begin{aligned}
{f_{QDR} }(\theta_l ,\varphi_l ,& \gamma_l ,\eta_l ) \\
=& {\left\| {{\xi^\#_l{\bm{a}}^H_l }{{\bm{U}}_N }} \right\|_F^2 } \\
=&( {\xi_{l1} ^* - \xi_{l2} ^*j_2 })C(\theta _l,\varphi _l)(\xi_{l1} ^* - \xi_{l2} ^*j_2 )^\ddag
 \end{aligned}
\end{equation}
where $C(\theta _l,\varphi _l)={{\bm{a}}^H_l }({\bm{U}}_{N 1}{\bm{U}}_{N 1}^H +{\bm{U}}_{N 2}^*{\bm{U}}_{N 2}^T){{\bm{a}}_l }, l=1,2,\cdots,L$. In general, when $0\leq \gamma _l<\pi/2$, $( {\xi_{l1} ^* - \xi_{l2} ^*j_2 })\neq0$. Thus $C(\theta _l,\varphi _l)=0$ means ${f_Q }  (\theta_l ,\varphi_l ,\gamma_l ,\eta_l ) =0$. The  spatial spectrum function of DOA estimation is given by
\begin{equation}\label{DOAQ}
{f_{QDR} }  (\theta_l ,\varphi_l )={{\bm{a}}^H_l }({\bm{U}}_{N 1}{\bm{U}}_{N 1}^H +{\bm{U}}_{N 2}^*{\bm{U}}_{N 2}^T){{\bm{a}}_l },
\end{equation}
$l=1,2,\cdots,L$. In general, substitute the result of the DOA estimation into (\ref{sfQMUSIC}), the polarization estimation can be obtained by searching the spectrum function in (\ref{sfQMUSIC}). However, the polarization parameter cannot be obtained. The reason is that after we take EVD of ${{\bm{R}}}$, the polarization information is contained in the eigenvalues, not the eigenvectors, i.e., the noise subspace ${{{\bm{U}}}_N }$ does not contain the polarization information. Thus the long vector (LV) method is reconsidered.

In practical situations, the theoretical array covariance matrices given in (\ref{DLCOV}) is unavailable and it can be estimated by
\begin{equation}\label{PRACOV2}
\hat{\bm{R}}_{LV}=\frac{1}{N}\{\overline{\bm{X}}{\kern 2pt}\overline{\bm{X}}^H\}.\\
\end{equation}

The EVD of covariance matrix can be expressed as
\begin{equation}\label{evdcs}
{{\bm{R}}_{LV}} = {{\hat{\bm{U}}}_S }{{\hat{\bm{\Sigma }}}_S }{\hat{\bm{U}}}_S ^H  + {{\hat{\bm{U}}}_N }{{\hat{\bm{\Sigma }}}_N }{\hat{\bm{U}}}_N ^H,
\end{equation}
where ${{\hat{\bm{U}}}_S }$ and ${{\hat{\bm{U}}}_N }$ are the signal and noise subspaces, respectively, the diagonal matrices ${{\hat{\bm{\Sigma }}}_S }$ and ${{\hat{\bm{\Sigma }}}_N }$ respectively contain their corresponding eigenvalues.

Since the DOA parameter has already been estimated, the manifold matrix $\overline{\bm{A}}\in \mathbb{C}^{2M\times L}$ in (\ref{outpa}) is merely the function of $(\gamma_l ,\eta_l )$. The spectrum function of polarization estimation is given by
\begin{equation}\label{PLQ}
{f_{QDR} }  (\gamma_l ,\eta_l )=\bm{a}^H_{Ql}{{\hat{\bm{U}}}_N }{{\hat{\bm{U}}}_N^H }\bm{a}_{Ql}, l=1,2,\cdots,L,
\end{equation}
where $\bm{a}_{Ql}$ is the steering vector belonging to $\overline{\bm{A}}$. Thus the pseudo code of the QDR-MUSIC can be summarized as {\bf{Algorithm 1}}.

\begin{algorithm}
  \caption{QDR-MUSIC  for Circular Signal}
  \begin{algorithmic}[1]
    \State Estimate $\hat{\bm{R}}$ according to (\ref{PRACOV});
 \State Construct the complex adjoint matrix ${\bm{R}^\sigma }$ of $\hat{\bm{R}}$ according to (\ref{evd});
 \State Take the EVD of  ${\bm{R}^\sigma }$ according to (\ref{cam});
 \State Construct the spectrum function ${f_{QDR} }  (\theta_l ,\varphi_l )$ according to (\ref{sfQMUSIC});
 \State Search (\ref{DOAQ}) to obtain the spectrum extremum $(\hat{\theta}_l ,\hat{\varphi}_l ), l=1,2,\cdots,L$;
 \State Estimate $\hat{\bm{R}}_{LV}$ according to (\ref{PRACOV2});
 \State Take the EVD of  ${\bm{R}_{LV}}$ according to (\ref{evdcs}) to acquire ${\hat{{\bm{U}}}_N}$;
 \State Estimate $({{\hat \gamma }_l}, {{\hat \eta }_l}), l=1,2,\cdots,L$ by searching (\ref{PLQ}).
  \end{algorithmic}
\end{algorithm}

 \vspace{-3em}
\subsection{Dimension-Reduced MUSIC Algorithm}

\subsubsection{DOA Estimation}

 Based on the principle of MUSIC algorithm, the subspace ${{\hat{\bm{U}}}_N }$ is orthogonal to  $ \mathrm{span}\begin{bmatrix}
{{{\bm{A}}}{{\bm{V}}_1}}\\
{{{\bm{A}}}{{\bm{V}}_2}}
\end{bmatrix}$, which is the space spanned by the array manifold matrix, then it holds that
\begin{equation}\label{sim}
{\begin{bmatrix}
   {{{\bm{A}}}{{\bm{V}}_1}}  \\
   {{{\bm{A}}}{{\bm{V}}_2}}  \\
\end{bmatrix}}^H {\hat{{\bm{U}}}_{N}} \\
= {\bm{V}}_1^H {\begin{bmatrix}
   {{{\bm{A}}}}  \\
   {{{\bm{A}}}{{\bm{V}}_2}{\bm{V}}_1^{ - 1}}  \\
\end{bmatrix}}^H {\hat{{\bm{U}}}_{N}} = {\bm{0}}.
\end{equation}
Partitioning ${\hat{{\bm{U}}}_{N }}$ into two block matrices $
{\hat{{\bm{U}}}_{N}} = \begin{bmatrix}
   {{\hat{{\bm{U}}}_{N 1 }}}  \\
   {{\hat{{\bm{U}}}_{N 2 }}}  \\
\end{bmatrix}$,
where $\hat{\bm{U}}_{N1}\in \mathbb{C}^{M\times (2M-L)}$ and $\hat{\bm{U}}_{N2}\in \mathbb{C}^{M\times (2M-L)}$ are noise subspaces, then we have
\begin{equation}
{\begin{bmatrix}
   {{{\bm{a}}_1}} & {{{\bm{a}}_2}} &  \cdots  & {{{\bm{a}}_L}}  \\
   {\frac{\xi_{21}}{\xi_{11}}{{\bm{a}}_1}} & {\frac{\xi_{22}}{\xi{12}}{{\bm{a}}_2}} &  \cdots  & {\frac{\xi_{2L}}{\xi_{1L}}{{\bm{a}}_L}}  \\
\end{bmatrix}}^H\\
\begin{bmatrix}
   {{\hat{{\bm{U}}}_{N 1 }}}  \\
   {{\hat{{\bm{U}}}_{N 2 }}}  \\
\end{bmatrix} = 0.
\end{equation}
Define a complex number as follows
\begin{equation}\label{rolde}
{\rho _l}{\exp{(i{\delta _l})}} = \frac{\xi_{2l}}{\xi_{1l}}.
\end{equation}
Based on (\ref{sim}), we can construct a spectrum function ${f_{DL }}(\theta ,\varphi ,\gamma ,\eta )$  as follows
\begin{equation}\label{DL}
\begin{aligned}
 {f_{DR }}&(\theta_l ,\varphi_l ,\gamma_l ,\eta_l )\\
  =& \left\| \begin{bmatrix}
   {{\bm{a}_l^H }} & {\rho_l {\exp{ (- i\delta )}}{\bm{a}_l^H }}  \\
\end{bmatrix}\begin{bmatrix}
   {{\hat{{\bm{U}}}_{N 1 }}}  \\
   {{\hat{{\bm{U}}}_{N 2}}}  \\
\end{bmatrix} \right\|_F^2 \\
= & {{\bm{a}_l}^H }{\hat{{\bm{U}}}_{N 1 }}\hat{{\bm{U}}}_{N 1 }^H {\bm{a}_l} + \rho_l {\exp{ (- i\delta) }}{\bm{a}_l^H }{\hat{{\bm{U}}}_{N 2}}\hat{{\bm{U}}}_{N 1}^H {\bm{a}_l} \\
 &+ \rho_l {\exp{(i\delta) }}{\bm{a}_l^H }{\hat{{\bm{U}}}_{N 1 }}\hat{{\bm{U}}}_{N 2 }^H {\bm{a}_l}+ {\rho_l ^2}{\bm{a}_l^H }{\hat{{\bm{U}}}_{N 2 }}\hat{{\bm{U}}}_{N 2 }^H {\bm{a}_l}. \\
 \end{aligned}
\end{equation}
By setting that the partial derivative of ${f_{DR }}(\theta_l ,\varphi_l ,\gamma_l ,\eta_l )$ with respect to $(\gamma_l ,\eta_l )$ equal to zero, we can obtain from (\ref{DL}) that
\begin{equation}\label{dff}
\begin{aligned}
{\exp{( - i\delta_l) }}{{\bm{a}}_l^H }{\hat{{\bm{U}}}_{N 2 }}\hat{{\bm{U}}}_{N 1 }^H {\bm{a}_l}& = {\exp{(i\delta_l) }}{{\bm{a}}_l^H }{\hat{{\bm{U}}}_{N 1}}\hat{{\bm{U}}}_{N 2}^H {\bm{a}_l}\\
 & =  - \rho_l {{\bm{a}}_l^H }{\hat{{\bm{U}}}_{N 2 }}\hat{{\bm{U}}}_{N 2}^H {\bm{a}_l},
\end{aligned}
\end{equation}
the proof is given in the supplemental materials. The spatial spectrum ${f_{DR }}(\theta_l ,\varphi_l ,\gamma_l ,\eta_l )$ in (\ref{DL}) minimizes according to $(\gamma_l ,\eta_l )$ when (\ref{dff}) takes the minus value. Substituting (\ref{dff}) into (\ref{DL}), we can simplify (\ref{DL}) as
\begin{equation}\label{rdl}
{f_{DR }}(\theta_l ,\varphi_l ,\rho_l ) = {{\bm{a}}_l^H }{\hat{{\bm{U}}}_{N 1 }}\hat{{\bm{U}}}_{N 1}^H {\bm{a}_l}- {\rho_l ^2}{{\bm{a}}_l^H }{\hat{{\bm{U}}}_{N 2 }}\hat{{\bm{U}}}_{N 2}^H {\bm{a}_l}.
\end{equation}
According to (\ref{dff}), ${\exp {(i\delta_l) }}{{\bm{a}}^H }{\hat{{\bm{U}}}_{N 1 }}\hat{{\bm{U}}}_{N 2 }^H {\bm{a}_l}$ is a real number, we have
\begin{equation}\label{deta}
{\exp{(i\delta) }}{{\bm{a}}^H }{\hat{{\bm{U}}}_{N 1 }}\hat{{\bm{U}}}_{N 2}^H {\bm{a}_l} = \sqrt {{{\left\| {{{\bm{a}}^H }{\hat{{\bm{U}}}_{N 1}}\hat{{\bm{U}}}_{N 2}^H {\bm{a}_l}} \right\|}_F^2}}.
\end{equation}
Then $\rho_l$ is given by
\begin{equation}\label{rou}
\rho_l  =  - \frac{{\sqrt {{{\left\| {{{\bm{a}}_l^H }{\hat{{\bm{U}}}_{N 1}}\hat{{\bm{U}}}_{N 2}^H {\bm{a}_l}} \right\|}_F^2}} }}{{{{\bm{a}}_l^H }{\hat{{\bm{U}}}_{N 2 }}\hat{{\bm{U}}}_{N 2}^H {\bm{a}_l}}}.
\end{equation}
Substituting (\ref{rou}) into (\ref{rdl}), we can simplify the spectrum function (\ref{rdl}) for the DOA estimation as
\begin{equation}\label{spectrumdl}
{f_{DR }}(\theta_l ,\varphi_l ) = {{\bm{a}}_l^H }{\hat{{\bm{U}}}_{N 1}}\hat{{\bm{U}}}_{N 1 }^H {\bm{a}_l}- \frac{{{{\left\| {{{\bm{a}}_l^H }{\hat{{\bm{U}}}_{N 1}}\hat{{\bm{U}}}_{N 2}^H {\bm{a}_l}} \right\|}_F^2}}}{{{{\bm{a}_l}^H }{\hat{{\bm{U}}}_{N 2}}\hat{{\bm{U}}}_{N 2}^H {\bm{a}_l}}}.
\end{equation}
Now the parameter search dimension of (\ref{spectrumdl}) has been reduced to two dimensions.

\subsubsection{Polarization Estimation}

The polarization estimation $({\hat{\gamma} _l},{\hat{\eta} _l})$ can be obtained when $L$ DOA estimations have been acquired. Based on (\ref{rolde}), (\ref{deta}) and (\ref{rou}), we have
\begin{equation}\label{polariz1}
c_l=  - \frac{{{{\left\| {{{{\hat{\bm{ a}}}}_l}^H {\hat{{\bm{U}}}_{N 1}}\hat{{\bm{U}}}_{N 2}^H {{{\hat{\bm{ a}}}}_l}} \right\|}_F^2}}}{{{{{\hat{\bm{ a}}}}_l}^H {\hat{{\bm{U}}}_{N 2 }}\hat{{\bm{U}}}_{N 2}^H {{{\hat{\bm{ a}}}}_l}}} \frac{1}{{{{{\hat{\bm{ a}}}}_l}^H {\hat{{\bm{U}}}_{N 1 }}\hat{{\bm{U}}}_{N 2}^H {{{\hat{\bm{ a}}}}_l}}} = \frac{\xi_{2l}}{\xi_{1l}}, \\
\end{equation}
where the steering vector ${\hat{\bm{ a}}_l}$ is constructed by $({\hat \theta _l},{\hat \varphi _l})$ which has already been estimated from (\ref{spectrumdl}) .

Substituting (\ref{components}) into (\ref{polariz1}), we have
\begin{equation}
\frac{{\sin {{\hat \theta }_l}\cos {{\hat \varphi }_l}\sin {\gamma _l}{\exp{(i {\eta _l})}} + \cos {{\hat \theta }_l}\cos {\gamma _l}}}{{\cos {{\hat \theta }_l}\cos {{\hat \varphi }_l}\sin {\gamma _l}{\exp {(i {\eta _l})}} - \sin {{\hat \theta }_l}\cos {\gamma _l}}} = {c_l},
\end{equation}
which can be simplified as
\begin{equation}
\tan ({\gamma _l}){\exp{(i {\eta _l})}} = \frac{{\cos {{\hat \theta }_l} + {c_l}\sin {{\hat \theta }_l}}}{{ - \sin {{\hat \theta }_l}\cos {{\hat \varphi }_l} + {c_l}\cos {{\hat \theta }_l}\cos {{\hat \varphi }_l}}}.
\end{equation}
Thus the polarization estimation of the $l$th signal is expressed as
\begin{equation}\label{estpol}
\begin{aligned}
 {{\hat \gamma }_l} &= \arctan \left( {\left\| {\frac{{\cos {{\hat \theta }_l} + {c_l}\sin {{\hat \theta }_l}}}{{ - \sin {{\hat \theta }_l}\cos {{\hat \varphi }_l} + {c_l}\cos {{\hat \theta }_l}\cos {{\hat \varphi }_l}}}} \right\|} \right), \\
 {{\hat \eta }_l}& = {\mathrm{arg}} \left( {\frac{{\cos {{\hat \theta }_l} + {c_l}\sin {{\hat \theta }_l}}}{{ - \sin {{\hat \theta }_l}\cos {{\hat \varphi }_l} + {c_l}\cos {{\hat \theta }_l}\cos {{\hat \varphi }_l}}}} \right). \\
  \end{aligned}
\end{equation}

Based on the method mentioned above, the DOA and polarization estimation are achieved. Thus the pseudo code of the DR-MUSIC can be summarized as {\bf{Algorithm 2}}.

\begin{algorithm}
  \caption{DR-MUSIC for Circular Signal}
  \begin{algorithmic}[1]
    \State Estimate $\hat{\bm{R}}_{LV}$ according to (\ref{PRACOV2});
 \State Take the EVD of  ${\bm{R}_{LV}}$ according to (\ref{evdcs});
 \State Construct the spectrum function ${f_Q }  (\theta_l ,\varphi_l)$ according to (\ref{spectrumdl});
 \State Search (\ref{spectrumdl}) to obtain the spectrum extremum $(\hat{\theta}_l ,\hat{\varphi}_l ), l=1,2,\cdots,L$.
 \State Calculate $c_l$ according to (\ref{polariz1});
 \State Estimate $({{\hat \gamma }_l}, {{\hat \eta }_l}), l=1,2,\cdots,L$ according to (\ref{estpol}).
  \end{algorithmic}
\end{algorithm}

\vspace{-1em}
The QDR-MUSIC and the DR-MUSIC can both be used for circular and non-circular signals. However, the property of non-circular signal, which can be utilized to improve the parameter estimation performance, has not been used in the QDR-MUSIC and the DR-MUSIC.

 \vspace{-1em}
\section{DOA and Polarization Estimation for Non-Circular Signals}
For a non-circular signal $s$, it holds that \cite{Abeida2006MUSIC}
\begin{equation}
\mathrm{E} \left[ {s(t)s(t)} \right] = \mu \exp{ ({i\beta })}\mathrm{E} \left[ {s(t){s^*}(t)} \right],
\end{equation}
in which $\beta $ is the non-circularity phase, $\mu$ is the non-circularity rate with $\mu  = 1$ for the maximal non-circularity rated signal and $0 < \mu  < 1$ for the common non-circularity rated signal.

For signal vector $\bm{S}\in\mathbb{C}^{L\times 1}$ consisting of $L$ independent components, its unconjugated covariance matrix is given by
\begin{equation}
\begin{aligned}
{{{\bm{R'}}}_{\bm{s}} } =&\mathrm{ E} \left[ {{\bm{S}}(t){{\bm{S}}^T }(t)} \right] \\
 =&{\rm diag} \{ {\mu _1}\exp{ ({i\beta_1 })}\mathrm{E} \left[ {{s_1}(t){s_1}^*(t)} \right], \\
&{\mu _2}\exp{ ({i\beta_2 })}\mathrm{E} \left[ {{s_2}(t){s_2}^*(t)} \right] , \ldots ,  \\
&{\mu _L}\exp{ ({i\beta_L })}\mathrm{E} \left[ {{s_L}(t){s_L}^*(t)} \right]\} \; \buildrel \Delta \over = {\bm{{P}{B}}}{{\bm{R}}_{\bm{s}} },\\
\end{aligned}
\end{equation}
where ${\bm{{P}}} = \mathrm{diag} \{ {\mu _1},{\mu _2}, \ldots ,{\mu _L}\} $ is a real valued diagonal matrix consisting of the non-circularity rates of $L$ signals.  ${\bm{{B}}} = \mathrm{diag} \{ \exp{ ({i\beta_1 })},\exp{ ({i\beta_2 })}, \ldots ,\exp{ ({i\beta_L })}\} $ is a diagonal matrix consisting of their non-circularity phases. For the maximal non-circularity rated signals, we have ${\bm{{P}}} = {\bm{I}_L}$.

%\begin{re}
%If the incident signals are non-circular signals, the property of non-circular signal can be used to improve the DOA estimation accuracy. In general, the maximal non-circularity rated signal is regarded as the non-circular signal.
%\end{re}
 \vspace{-1em}
\subsection{DOA Estimation}
In order to utilize the information contained in the unconjugated covariance matrix of the non-circular signals, two quaternion vectors are constructed as
\begin{equation}\label{qra}
{\bm{y}}(t) = {{\bm{x}}_1}(t) + j_2 {{\bm{x}}_2}(t),{\bm{z}}(t) =  {\bm{x}}_1^*(t) + j_2 {\bm{x}}_2^*(t).
\end{equation}
Substituting (\ref{outpa}) into (\ref{qra}), we have
\begin{equation}\label{oy}
\begin{aligned}
\bm{y}(t) &= {{\bm{A}}^ * }\left( { {\bm{V}}_1^{} + j_2 {\bm{V}}_2^{}} \right){\bm{s}}(t) + \left( { {\bm{n}}_1^{}(t) + j_2 {\bm{n}}_2^{}(t)} \right)\\ &\buildrel \Delta \over = {{\bm{A}}^ * }{{\bm{V}}_y}{\bm{s}}(t) + {{\bm{n}}_y}(t),
\end{aligned}
\end{equation}
\begin{equation}\label{oz}
\begin{aligned}
{\bm{z}}(t) &= {\bm{A}}\left( { {\bm{V}}_1^* + j_2 {\bm{V}}_2^*} \right){{\bm{s}}^*}(t) + \left( { {\bm{n}}_1^*(t) + j_2 {\bm{n}}_2^*(t)} \right)\\ &\buildrel \Delta \over = {\bm{A}}{{\bm{V}}_z}{{\bm{s}}^*}(t) + {{\bm{n}}_z}(t).
\end{aligned}
\end{equation}
where the diagonal entries of the quaternion matrices ${{\bm{V}}_y}$ and ${{\bm{V}}_z}$ are ${v_{yl}}$ and ${v_{zl}}, l=1,2,\cdots, L$, respectively, which can be expressed as
\begin{equation}\label{cvd}
{v_{yl}} =  {\xi_{1l} } + j_2 {\xi_{2l} },{v_{zl}} =  \xi_{1l} ^* + j_2 \xi_{2l} ^*.
\end{equation}

The quaternion vector ${\bm{w}}(t)\in \mathbb{H}^{2M\times1}$ is constructed as
\begin{equation}
\begin{aligned}
{\bm{w}}(t) =\begin{bmatrix}
{{\bm{y}}(t)} \\
{{\bm{z}}(t)}\\
\end{bmatrix} &= \begin{bmatrix}
{{{\bm{x}}_1}(t)} \\
{{\bm{x}}_1^*(t)}\\
\end{bmatrix}+ j_2\begin{bmatrix}
{{{\bm{x}}_2}(t)} \\
{{\bm{x}}_2^*(t)}\\
\end{bmatrix}\\
&={\bm{w}_1}(t)+j_2{\bm{w}_2}(t).
\end{aligned}
\end{equation}
The quaternion covariance matrix ${{\bm{R}}_w}\in \mathbb{H}^{2M\times2M}$ of the extended vector ${\bm{w}}(t)$ is expressed as
\begin{equation}\label{cov}
{{\bm{R}}_w} = \mathbb{E }\left[ {{\bm{w}}(t){{\bm{w}}^\ddag }(t)} \right]=\begin{bmatrix}
{\mathbb{E} \left[ {{\bm{y}}(t){{\bm{y}}^\ddag }(t)} \right]}& {\mathbb{E} \left[ {{\bm{y}}(t){{\bm{z}}^\ddag }(t)} \right]}\\
{\mathbb{E} \left[ {{\bm{z}}(t){{\bm{y}}^\ddag }(t)} \right]} & {\mathbb{E} \left[ {{\bm{z}}(t){{\bm{z}}^\ddag }(t)} \right]}
\end{bmatrix}.
\end{equation}
According to (\ref{oy}) and (\ref{oz}), we have
\begin{equation}\label{yy}
\begin{aligned}
\mathbb{E} \left[ {{\bm{y}}(t){{\bm{y}}^\ddag }(t)} \right] &= {{\bm{A}}^*}{{\bm{V}}_y}{{\bm{R}}_S }{\bm{V}}_y^\ddag {{\bm{A}}^T } + E \left[ {{{\bm{n}}_y}(t){{\bm{n}}_y}^\ddag (t)} \right]\\
& = {{\bm{A}}^*}{{\bm{V}}_y}{{\bm{R}}_{\bm{s}} }{\bm{V}}_y^\ddag {{\bm{A}}^T } + 2\sigma _n ^2{\bm{I}_M},
\end{aligned}
\end{equation}
\begin{equation}\label{yz}
\begin{aligned}
\mathbb{E} \left[ {{\bm{y}}(t){{\bm{z}}^\ddag }(t)} \right]& = {{\bm{A}}^*}{{\bm{V}}_y}{{\bm{R'}}_S }{\bm{V}}_z^\ddag {{\bm{A}}^H } + \mathbb{E} \left[ {{{\bm{n}}_y}(t){{\bm{n}}_z}^\ddag (t)} \right]\\
& = {{\bm{A}}^*}{{\bm{V}}_y}{{\bm{R'}}_{\bm{s}} }{\bm{V}}_z^\ddag {{\bm{A}}^H },
\end{aligned}
\end{equation}
\begin{equation}\label{zz}
\begin{aligned}
\mathbb{E} \left[ {{\bm{z}}(t){{\bm{z}}^\ddag }(t)} \right]& = {\bm{A}}{{\bm{V}}_z}{{\bm{R}}_S }{\bm{V}}_z^\ddag {{\bm{A}}^H } + \mathbb{E} \left[ {{{\bm{n}}_z}(t){{\bm{n}}_z}^\ddag (t)} \right] \\
&= {\bm{A}}{{\bm{V}}_z}{{\bm{R}}_{\bm{s}} }{\bm{V}}_z^\ddag {{\bm{A}}^H } + 2\sigma _n ^2{\bm{I}_M}.
\end{aligned}
\end{equation}
Substituting (\ref{yy}), (\ref{yz}) and (\ref{zz}) into (\ref{cov}) and using the relationship $\;{{{\bm{R'}}}_{\bm{s}} }= {\bm{{P}{B}}}{{\bm{R}}_{\bm{s}} }$,
 we can rewrite ${{\bm{R}}_w}$ as
\begin{equation}\label{NCEVDD}
\begin{aligned}
 {{\bm{R}}_w}=&\begin{bmatrix}
{{{\bm{A}}^*}{{\bm{V}}_y}{{\bm{R}}_{\bm{s}} }{\bm{V}}_y^\ddag {{\bm{A}}^T }}& {{{\bm{A}}^*}{{\bm{V}}_y}{{ \bm{B}}}{{\bm{R}}_{\bm{s}} }{\bm{V}}_z^\ddag {{\bm{A}}^H }}\\
{{\bm{A}}{{\bm{V}}_z}{{\bm{{ B}}}^*}{{\bm{R}}_{\bm{s}} }{\bm{V}}_y^\ddag {{\bm{A}}^T }} & {{\bm{A}}{{\bm{V}}_z}{{\bm{R}}_{\bm{s}} }{\bm{V}}_z^\ddag {{\bm{A}}^H }}
\end{bmatrix} \\
&+ 2\sigma _n ^2{\bm{I}_{2M}} \\
  =& \begin{bmatrix}
   {{{\bm{A}}^*}{{\bm{V}}_y}} \\
   {{\bm{A}}{{\bm{V}}_z}{{\bm{{ B}}}^*}}  \\
\end{bmatrix} \bm{R}_{\bm{s}}{\begin{bmatrix}
   {{{\bm{A}}^*}{{\bm{V}}_y}}\\
   {{\bm{A}}{{\bm{V}}_z}{{\bm{{ B}}}^*}}   \\
\end{bmatrix} }^\ddag  + 2\sigma _n ^2{\bm{I}_{2M}} \\
  \buildrel \Delta \over = & \hat{\bm{ A}}{{\bm{ R}}_{\bm{s}} }{\hat{{\bm{ A}}}^\ddag } + 2\sigma _n ^2{\bm{I}_{2M}}. \\
 \end{aligned}
\end{equation}

In practical situations, the theoretical array covariance matrices given in (\ref{cov}) is unavailable and it is usually estimated by
\begin{equation}\label{PRACOV3}
\begin{aligned}
\hat{\bm{R}}_w&=\frac{1}{N}\{\bm{W}\bm{W}^\ddag\}=\frac{1}{N}({\bm{W}_1}+j_2{\bm{W}_2})({\bm{W}_1}+j_2{\bm{W}_2})^\ddag\\
&=\frac{1}{N}[\bm{W}_1\bm{W}_1^H+\bm{W}_2^*\bm{W}_2^T+j_2(\bm{W}_2\bm{W}_1^H-\bm{W}_1^*\bm{W}_2^T)]\\
&\buildrel \Delta \over=\bm{R}_{w1}+j_2\bm{R}_{w2},
\end{aligned}
\end{equation}
where $\bm{W}\in \mathbb{H}^{2M\times N}$, $\bm{W}_1\in \mathbb{C}^{2M\times N}$ and $\bm{W}_2\in \mathbb{C}^{2M\times N}$ are the snapshot data matrices constructed by $\bm{w}(t)$, $\bm{w}_1(t)$ and $\bm{w}_2(t)$, respectively, $t=t_1,\dots,t_N$.

Similar to (\ref{cam}), the EVD of complex Hermite adjoint matrix ${{\bm{R}}^\sigma_w }$ can be obtained as well, thus the EVD of ${{\bm{R}}_w}$ can be expressed as
\begin{equation}\label{wevd}
{{\bm{R}}_w} = {{\overline{\bm{U}}}_S }{{\overline{\bm{\Sigma }}}_S }{\overline{\bm{U}}}_S ^\ddag  + {{\overline{\bm{U}}}_N }{{\overline{\bm{\Sigma }}}_N }{\overline{\bm{U}}}_N ^\ddag.
\end{equation}
Based on the orthogonality of subspace ${{\overline{\bm{U}}}_S }\in \mathbb{H}^{2M\times L}$ and ${{\overline{\bm{U}}}_N \in \mathbb{H}^{2M\times(2M-L)}}$, it holds that
\begin{equation}\label{orth}
{\hat{\bm{ A}}^\ddag }{{\overline{\bm{U}}}_N } = {\bm{0}}.
\end{equation}

Let us partition the noise subspace into two block matrices $
{{\overline{\bm{U}}}_N } =\begin{bmatrix}
{{{\overline{\bm{U}}}_{N 1}}}\\
{{{\overline{\bm{U}}}_{N 2}}}
\end{bmatrix}$.
Then the spectrum function of quaternion non-circular MUSIC (QNC-MUSIC) is constructed as
\begin{equation}\label{QNC}
\begin{aligned}
 f_{QNC}(\theta_l ,& \varphi_l ,\gamma_l ,\eta_l )\\
  =& {\left\| {{\begin{bmatrix}
   {{{\bm{a}}^*_l}v_{yl}} & {{\bm{a}_l}{v_{zl}}{\exp {( - i \beta_l) }}}  \\
\end{bmatrix}^\ddag }\begin{bmatrix}
{{{\overline{\bm{U}}}_{N 1}}}\\
{{{\overline{\bm{U}}}_{N 2}}}
\end{bmatrix}} \right\|_F^2 }. \\
 \end{aligned}
\end{equation}

We define three quaternion numbers $
{{\Pi }_1} = {{\bm{a}}^T }{{\overline{\bm{U}}}_{N 1}}{\overline{\bm{U}}}_{N 1}^\ddag {{\bm{a}}^*}$, ${{\Pi}_2} = {{\bm{a}}^H }{{\overline{\bm{U}}}_{N 2}}{\overline{\bm{U}}}_{N 2}^\ddag {\bm{a}}$ and ${{\Pi }_3} = {{\bm{a}}^H }{{\overline{\bm{U}}}_{N 2}}{\overline{\bm{U}}}_{N 1}^\ddag {{\bm{a}}^*}$. Then (\ref{QNC}) can be written as
\begin{equation}\label{QNC1}
\begin{aligned}
 f_{QNC} (\theta_l ,& \varphi_l ,\gamma_l ,\eta_l ) \\
 =& \left( {{{\left| {{\xi_{l1} }} \right|}^2} + {{\left| {{\xi_{l2} }} \right|}^2} + 2{{\rm real}} \left\{ {i {\xi_{l1} }\xi_{l2} ^*} \right\}} \right){{\Pi }_1} \\
 &+ \left( {{{\left| {{\xi_{l1} }} \right|}^2} + {{\left| {{\xi_{l2}}} \right|}^2} - 2{{\rm real}} \left\{ {i {\xi_{l1} }\xi_{l2} ^*} \right\}} \right){{\Pi }_2} \\
  &+ {\exp{(i \beta_l) }}\left( {{\xi_{l1} }^2 + {\xi_{l2} }^2} \right){\Pi }_3^* \\
 & + {\exp{( - i \beta_l) }}\left( |\xi_{l1}^*|^2+ |\xi_{l2}^*|^2 \right){{\Pi}_3}. \\
 \end{aligned}
\end{equation}
Let the partial derivative of $f_{QNC}(\theta_l ,\varphi_l ,\gamma_l ,\eta_l )$ with respect to $\beta_l$ equal to zero, we can get
\begin{equation}\label{der}
\begin{aligned}
\frac{{\partial {f_{QNC}(\theta_l ,\varphi_l ,\gamma_l ,\eta_l )}}}{{\partial \beta_l }}=& i {\exp{(i \beta_l) }}\left( {{\xi_{l1} }^2 + {\xi_{l2} }^2} \right){\Pi }_3^*\\
&- i {\exp{ (- i \beta_l) }}\left( |\xi_{l1}^*|^2+ |\xi_{l2}^*|^2 \right){{\Pi }_3}\\
 = &0.
\end{aligned}
\end{equation}
The non-circularity phase after calculating (\ref{der}) is given by
\begin{equation}\label{beta}
{\exp {(i \beta_l) }} =  - \frac{{\left( |\xi_{l1}^*|^2+ |\xi_{l2}^*|^2 \right){{\Pi }_3}}}{{\left| {\left( |\xi_{l1}|^2+ |\xi_{l2}|^2 \right){\Pi }_3^*} \right|}}.
\end{equation}
Substituting (\ref{beta}) into (\ref{QNC1}), we can obtain the simplified spectrum function for DOA and polarization estimation of non-circular signals as
\begin{equation}\label{QNC2}
\begin{aligned}
 f_{QNC}& (\theta_l ,\varphi_l ,\gamma_l ,\eta_l )\\
 = & \left( {{{\left| {{\xi_{l1} }} \right|}^2} + {{\left| {{\xi_{l2} }} \right|}^2} + 2{{\rm real}} \left\{ {i {\xi_{l1} }\xi_{l2}^*} \right\}} \right){{\Pi }_1} \\
 & + \left( {{{\left| {{\xi_{l1} }} \right|}^2} + {{\left| {{\xi_{l2} }} \right|}^2} - 2{{\rm real}} \left\{ {i {\xi_{l1} }\xi_{l2} ^*} \right\}} \right){{\Pi }_2} \\
 & - 2\left| {\left( {|\xi_{l1}|^2+ |\xi_{l2}|^2 } \right){\Pi}_3^*} \right|. \\
 \end{aligned}
\end{equation}
However, the DOAs of the incident signals can be estimated by the proposed algorithm, but the polarizations of the incident signals cannot be estimated by the proposed algorithm.

One possible reason is that the constraint condition of the polarization parameters is relatively weak. The DOA parameters are contained not only in steering vector ${\bm{a}_l}$, but also in ${\xi_{1l} }$ and ${\xi_{2l} }$, The polarization parameters are merely contained in ${\xi_{1l} }$ and ${\xi_{2l} }$. Thus the condition that the polarization estimation with multiple solutions may happen by searching (\ref{QNC2}). Another possible reason is that the polarization parameters are presented as a quaternion (this can be found in (\ref{cvd})). After we take the EVD of ${{\bm{R}}_w}$, a major part of the polarization information is contained in the eigenvalues, not the eigenvectors, i.e., the noise subspace ${{\overline{\bm{U}}}_N }$ sparingly contains the polarization information. Thus the polarization parameter cannot be estimated using (\ref{orth}).
\begin{algorithm}
  \caption{QNC-MUSIC for Non-Circular Signals}
  \begin{algorithmic}[1]
    \State Estimate $\hat{\bm{R}}_w$ according to (\ref{PRACOV3});
 \State Construct the complex adjoint matrix ${\bm{R}_w^\sigma }$ of $\hat{\bm{R}}_w$ according to (\ref{evd});
 \State Take the EVD of  ${\bm{R}^\sigma_w }$ according to (\ref{wevd});
 \State Construct the spectrum function ${f_{QNC} }  (\theta_l ,\varphi_l ,\gamma_l ,\eta_l )$ according to (\ref{QNC2});
 \State Search (\ref{QNC2}) to obtain the spectrum extremum $(\overline{\theta}_l ,\overline{\varphi}_l  ), l=1,2,\cdots,L$.
 \State Estimate $\hat{\bm{R}}_{LV}$ according to (\ref{PRACOV2});
 \State Take the EVD of  ${\bm{R}_{LV}}$ according to (\ref{evdcs});
 \State Partition ${\hat{{\bm{U}}}}$ into ${\hat{{\bm{U}}}_{N 1}}$ and ${\hat{{\bm{U}}}_{N 2}}$;
  \State Calculate $c_l$ according to (\ref{polariz1});
 \State Estimate $({{\overline{ \gamma }}_l}, {{\overline{\eta} }_l}), l=1,2,\cdots,L$ according to (\ref{estpol}).
  \end{algorithmic}
\end{algorithm}
\vspace{-1em}
\vspace{-1em}
\subsection{Polarization Estimation}
The noise subspace containing the polarization information has been estimated from (\ref{evdcs}), thus the polarization estimation $({\hat{\gamma} _l},{\hat{\eta} _l})$ can be obtained as well when $L$ DOAs of non-circular signals have been acquired from (\ref{QNC2}). The steering vector ${\overline{\bm{ a}}_l}$ is constructed by $({\overline\theta _l},{\overline \varphi _l})$. Based on (\ref{polariz1}), we have
\begin{equation}\label{polariz}
\begin{aligned}
d_l& =  - \frac{{{{\left\| {{{{\overline{\bm{ a}}}}_l}^H {\hat{{\bm{U}}}_{N 1}}\hat{{\bm{U}}}_{N 2}^H {{{\overline{\bm{ a}}}}_l}} \right\|}_F^2}}}{{{{{\overline{\bm{ a}}}}_l}^H {\hat{{\bm{U}}}_{N 2 }}\hat{{\bm{U}}}_{N 2}^H {{{\overline{\bm{ a}}}}_l}}{{{{{\overline{\bm{ a}}}}_l}^H {\hat{{\bm{U}}}_{N 1 }}\hat{{\bm{U}}}_{N 2}^H {{{\overline{\bm{ a}}}}_l}}}}. \\
 \end{aligned}
\end{equation}
Then the polarization estimation ${{\overline \gamma }_l}, {{\overline\eta }_l}$ of the $l$th non-circular signal can be achieved as the step 5 and 6 in \textbf{Algorithm 2}. Thus the DOA and polarization estimation for non-circular signals are achieved. The pseudo code of the QNC-MUSIC can be summarized as {\bf{Algorithm 3}}.

\section{Computational Complexity Analysis}
In this section, we analyze the computational complexity of the QNC-MUSIC and the DR-MUSIC and compare their computational complexity of them to those of the classical LV-MUSIC and the QDR-MUSIC. The computational complexity of the MUSIC-based algorithm concentrates on the estimation of the covariance matrix, the EVD of the covariance matrix, the spectrum search of the DOA parameter $(\theta, \varphi)$ and the spectrum search of the polarization parameter $(\gamma, \eta)$, respectively. We compare the four algorithms in the four parts mentioned above.

(1) \textit{The computational complexity of the estimation of the covariance matrix}

For the classical LV-MUSIC, the dimension of the received data matrix is $2M\times N$, thus $4{M^2} N$ flops are required for the calculation of the covariance matrix. For the QDR-MUSIC and the QNC-MUSIC, the complex adjoint matrices (\ref{PRACOV}) and (\ref{PRACOV3}) of the quaternion covariance matrix $\bm{R}$ and $\bm{R}_w$ require $4{M^2} N$ flops to be calculated, respectively. For the DR-MUSIC, the dimension of the received data matrix is identical with that of the LV-MUSIC, thus $4{M^2} N$ flops are required for the calculation of the covariance matrix as well.

(2) \textit{The computational complexity of the EVD on the covariance matrix}

For the LV-MUSIC and the DR-MUSIC, the EVD of the  covariance matrix are using the fast subspace decomposition (FSD) technique \cite{Xu1994Fast}, and their complexity is given by $4M^2(L+2)$ flops. For the QDR-MUSIC and the QNC-MUSIC using the FSD technique, the computational complexity of the EVD of the complex adjoint matrices $\bm{R}^\sigma\in \mathbb{C}^{2M\times 2M}$ and $\bm{R}^\sigma_w\in \mathbb{C}^{4M\times 2M}$ are $4M^2(L+2)$ and $16M^2(L+2)$ flops, respectively.

(3) \textit{The computational complexity of the spectrum search of the DOA parameter $(\theta, \varphi)$}

Let $J_1$, $J_2$, $J_3$ and $J_4$ stand for the number of spectral points for parameter $\theta$, $\varphi$, $\gamma$ and $\eta$, respectively. For the LV-MUSIC, it has to compute $\|\bm{a}_{LV}^H(\theta ,\varphi ,\gamma ,\eta )\hat{\bm{U}}_{N}\|^2$ for each spectral point, the spectral search step costs $J_1J_2J_3J_4(2M+1)(2M-L)$ flops, where $\bm{a}_{LV}(\theta ,\varphi ,\gamma ,\eta )\in \mathbb{C}^{2M\times 1}$ is the steering vector belonging to the LV-MUSIC, and $\hat{\bm{U}}_{N}\in \mathbb{C}^{2M\times (2M-L)}$ is the noise subspace. For the QDR-MUSIC, it has to compute ${{\bm{a}}^H(\theta ,\varphi) }({\bm{U}}_{N 1}{\bm{U}}_{N 1}^H +{\bm{U}}_{N 2}^*{\bm{U}}_{N 2}^T){{\bm{a}}(\theta ,\varphi) }$ for each spectral point, where $\bm{a}(\theta ,\varphi)\in \mathbb{C}^{M\times 1}$ is the steering vector, and $\bm{U}_{N1}, \bm{U}_{N2}\in \mathbb{C}^{M\times (M-L)}$ are the noise subspaces, then the spectral search step costs $2J_1J_2(M+1)(M-L)$. For the QNC-MUSIC, it has to compute ${{\Pi }_1} = {{\bm{a}}^T (\theta ,\varphi )}{{\overline{\bm{U}}}_{N 1}}{\overline{\bm{U}}}_{N 1}^\ddag {{\bm{a}}^*(\theta ,\varphi )}$, ${{\Pi}_2} = {{\bm{a}}^H (\theta ,\varphi )}{{\overline{\bm{U}}}_{N 2}}{\overline{\bm{U}}}_{N 2}^\ddag {\bm{a}(\theta ,\varphi )}$ and ${{\Pi }_3}= {{\bm{a}}^H(\theta ,\varphi ) }{{\overline{\bm{U}}}_{N 2}}{\overline{\bm{U}}}_{N 1}^\ddag {{\bm{a}}^*(\theta ,\varphi )}$ for each spectral point, where $\bm{a}(\theta ,\varphi )\in \mathbb{C}^{M\times 1}$ is the steering vector, $\bm{U}_{N1}\in \mathbb{H}^{M\times (2M-L)}$ and $\bm{U}_{N2}\in \mathbb{H}^{M\times (2M-L)}$ are the quaternion matrices. Since the intermediate results of the calculation of $\Pi_1$ and $\Pi_2$ can be used for the calculation of $\Pi_3$, the spectral search step costs $4J_1J_2(M+1)(2M-L)+2J_1J_2(2M-L)$ flops. For the DR-MUSIC, it has to compute ${{\bm{a}}^H (\theta ,\varphi )}{\hat{{\bm{U}}}_{N 1}}\hat{{\bm{U}}}_{N 1 }^H {\bm{a}(\theta ,\varphi )}$, ${{\bm{a}}^H (\theta ,\varphi )}{\hat{{\bm{U}}}_{N 2}}\hat{{\bm{U}}}_{N 2 }^H {\bm{a}(\theta ,\varphi )}$ and ${{\bm{a}}^H(\theta ,\varphi ) }{\hat{{\bm{U}}}_{N 1}}\hat{{\bm{U}}}_{N 2 }^H {\bm{a}(\theta ,\varphi )}$ for each spectral point, where $\bm{a}(\theta ,\varphi )\in \mathbb{C}^{M\times 1}$ is the steering vector, $\hat{\bm{U}}_{N1}\in \mathbb{C}^{M\times (2M-L)}$ and $\hat{\bm{U}}_{N2}\in \mathbb{C}^{M\times (2M-L)}$ are complex matrices. The spectral search step costs $2J_1J_2(M+1)(2M-L)+J_1J_2(2M-L)$ flops, which is similar as the calculation process of the QNC-MUSIC.

(4) \textit{The computational complexity of the spectrum search of the polarization parameter $(\gamma ,\eta )$}

For the LV-MUSIC,  the four dimensional (4-D) parameter search is used for jointly estimating $\theta$, $\varphi$, $\gamma$ and $\eta$, thus the total complexity for the DOA and polarization estimation is $J_1J_2J_3J_4(2M+1)(2M-L)$ flops. For the QDR-MUSIC, $\hat{\bm{U}}_{N}\in \mathbb{C}^{2M\times (2M-L)}$ in (\ref{evdcs}) is needed to be calculated, thus the EVD of (\ref{evdcs}) costs $4{M^2} N+4M^2(L+2)$ flops. For each incident signal, it has to compute $\|\bm{a}_{Ql}^H(\gamma ,\eta )\hat{\bm{U}}_{N}\|^2$ for each spectral point, the spectral search step costs $J_3J_4(2M+1)(2M-L)$ flops, where $\bm{a}_{Ql}\in \mathbb{C}^{2M\times 1}$ is the steering vector of the $l$th signal. The total computational complexity of the QDR-MUSIC is $J_3J_4L(2M+1)(2M-L)+4{M^2} N+4M^2(L+2)$. For the QNC-MUSIC, $\hat{\bm{U}}_{N1}\in \mathbb{C}^{M\times (2M-L)}$ and $\hat{\bm{U}}_{N2}\in \mathbb{C}^{M\times (2M-L)}$ in (\ref{polariz}) are needed to be calculated, thus the EVD of (\ref{evdcs}) has to be taken. According to (\ref{estpol}), it can be seen that an analytical solution exists for the polarization parameter $(\gamma ,\eta )$, the spectrum search is not needed. Thus the computational complexity for the polarization parameter $(\gamma ,\eta )$ estimation is $4{M^2} N+4M^2(L+2)$ flops. For the DR-MUSIC, the spectrum search is not needed for the polarization parameter $(\gamma ,\eta )$ estimation, which is similar as that of the QNC-MUSIC.

We can summarize the computational complexity of the four algorithms as follows. The computational complexity of the LV-MUSIC, the QDR-MUSIC, QNC-MUSIC and DR-MUSIC are given by $C_{LV}=J_1J_2J_3J_4(2M+1)(2M-L)+4M^2(N+L+2)$ flops, $C_{QDR}=2J_1J_2(M+1)(M-L)+J_3J_4L(2M+1)(2M-L)+8M^2(N+L+2)$ flops, $C_{QNC}=4J_1J_2(M+1)(2M-L)+2J_1J_2(2M-L)+20M^2(L+2)+8M^2N$ flops. and $C_{DR}=2J_1J_2(2M+1)(2M-L)+J_1J_2(2M-L)+4M^2(N+L+2)$ flops, respectively.

The computational complexity of the LV-MUSIC $(C_{LV}\rightarrow 4M^2J_1J_2J_3J_4)$, QDR-MUSIC $(C_{QDR}\rightarrow 2M^2J_1J_2+4LM^2J_3J_4)$, QNC-MUSIC $(C_{QNC}\rightarrow 8M^2J_1J_2)$ and the DR-MUSIC $(C_{DR}\rightarrow 8M^2J_1J_2)$, respectively, as $(M\rightarrow\infty)$.  When the source number increases, the computational complexity of the QDR-MUSIC would increase greatly. Obviously, in terms of implementation, the LV-MUSIC and the QDR-MUSIC algorithms are significantly more complicated than that of the QNC-MUSIC and the DR-MUSIC algorithms in the context of the massive MIMO systems.

\vspace{-1em}
\section{Stochastic Cram\'{e}r-Rao Bound for Non-Circlar Sources}
The stochastic CRB \cite{Stoica1990Performance, Stoica2001The, delmas2004stochastic,hassen2011doa} can be achieved asymptotically (in the number of measurements) by the stochastic maximum likelihood (ML) method as the (asymptotic) covariance matrix of the ML estimator. The parameter estimation can achieve a higher estimation accuracy and can resolve up to twice as many signal sources compared to the traditional methods for arbitrary signals via exploiting the structure of second-order non-circular (NC) signals \cite{delmas2004stochastic, hassen2011doa}.

In the following, we will be concerned with the signal model
\begin{equation}\label{crboutpa}
\overline{\bm{x}}\left( t \right)=\overline{\bm{A}}(\bm{\vartheta})\bm{S}(t)+\bm{N}(t),\quad t=1,...,N,
\end{equation}
where $\overline{\bm{A}}(\bm{\vartheta})\in\mathbb{C}^{2M\times L}$ is the steering matrix. The parameter $\bm{\vartheta}^{T}=[\bm{\theta}^{T}, \bm{\psi}^{T}, \bm{\gamma}^{T}, \bm{\eta}^{T}]^{T} \in \mathbb{R}^{4L\times 1}$ and $\bm{\theta}$, $\bm{\psi}$, $\bm{\gamma}$, $\bm{\eta}$ are all in $\mathbb{R}^{L}$. Non-Circlar Sources $\bm{S}(t)$ with $\bm{R}_{\bm{s}}={\rm{E}}\{\bm{S}(t)\bm{S}(t)^{H}\}$ and $\bm{R}_{\bm{s}}^{\prime}={\rm{E}}\{\bm{S}(t)\bm{S}(t)^{T}\}$, circular complex noise $\bm{N}(t)$ with unconjugated covariance matrix ${\rm{E}}\{\bm{N}(t)\bm{N}(t)^{T}\}=\bm{0}$. Thus we have
\begin{equation}
\bm{R}_{\overline{\bm{x}}}=\overline{\bm{A}}\bm{R}_{\bm{s}}\overline{\bm{A}}^{H}+\sigma_{n}^{2}\bm{I}_{2M}\,\;\textrm{and}\,\;\bm{R}_{\overline{\bm{x}}}^{\prime}=\overline{\bm{A}}
\bm{R}_{\bm{s}}^{\prime}\overline{\bm{A}}^{T}.
\end{equation}

If no a priori information is available, $(\bm{R}_{\overline{\bm{x}}},\bm{R}_{\overline{\bm{x}}}^{\prime})$ is generically parameterized by the real unknown parameter vector $\bm{\Theta}:=[\bm{\vartheta}^{T},\bar{\bm{\alpha}}^{T},\sigma_{n}]^{T}$ and
\begin{align}
\bar{\bm{\alpha}}=&\Big[\Big(\textrm{Re}\big([\bm{R}_{\bm{s}}]_{\bar{i},\bar{j}}\big), \textrm{Im}\big([\bm{R}_{\bm{s}}]_{\bar{i},\bar{j}}\big),\textrm{Re}\big([\bm{R}_{\bm{s}}^{\prime}]_{\bar{i},\bar{j}}\big),\nonumber\\ &\textrm{Im}\big([\bm{R}_{\bm{s}}^{\prime}]_{\bar{i},\bar{j}}\big)\Big)_{1\leq \bar{j}<\bar{i}\leq L},\Big(\big([\bm{R}_{\bm{s}}]_{\bar{i},\bar{i}}\big),\textrm{Re}\big([\bm{R}_{\bm{s}}^{\prime}]_{\bar{i},\bar{i}}\big),\nonumber\\
&\textrm{Im}\big([\bm{R}_{\bm{s}}^{\prime}]_{\bar{i},\bar{i}}\big)\Big)_{\bar{i}=1,...,L}\Big]^{T}\in \mathbb{R}^{\big(L^{2}+L(L+1)\big)\times 1}.
\end{align}

The probability density function of $\overline{\bm{x}}\left( t \right)$ is expressed as a function $\tilde{\bm{x}}(t):=\Big[
\begin{array}{c} \overline{\bm{x}}\left( t \right)\\ \overline{\bm{x}}^{*}\left( t \right)\\ \end{array}
\Big] \in \mathbb{C}^{4M}$ in the case of uniform white noise.
\begin{equation}
p(\tilde{\bm{x}}(t))=(\pi)^{-2M}\big|\bm{R}_{\tilde{\bm{x}}}\big|^{-\frac{1}{2}}\textrm{exp}\Big[-\frac{1}{2}\tilde{\bm{x}}^{H}
\bm{R}_{\tilde{\bm{x}}}^{-1}\tilde{\bm{x}}\Big],
\label{eq:1pdf}
\end{equation}
where
\begin{equation}
\bm{R}_{\tilde{\bm{x}}}={\rm{E}}\{\tilde{\bm{x}}\tilde{\bm{x}}^{H}\}=\tilde{\bm{A}}\bm{R}_{\tilde{s}}\tilde{\bm{A}}^{H}+\bm{R}_{\tilde{n}}
\end{equation}
with
\begin{equation}
\bm{R}_{\tilde{\bm{s}}}= \left[
\begin{array}{cc}
\bm{R}_{\bm{s}} & \bm{R}_{\bm{s}}^{\prime} \\
\bm{R}_{\bm{s}}^{\prime*} & \bm{R}_{\bm{s}}^{*} \\
\end{array} \right]
\quad
\tilde{\bm{A}}= \left[
\begin{array}{cc}
\overline{\bm{A}} & \bm{0}_{2M\times L} \\
\bm{0}_{2M\times L} & \overline{\bm{A}}^{*} \\
\end{array} \right]
\end{equation}
and
\begin{equation}
\bm{R}_{\tilde{\bm{n}}}= \left[
\begin{array}{cc}
\sigma_{n}^{2}\bm{I}_{2M} & \bm{0}_{2M\times 2M} \\
\bm{0}_{2M\times 2M} & \sigma_{n}^{2}\bm{I}_{2M}^{*} \\
\end{array} \right]=\sigma_{n}^{2}\bm{I}_{4M}.
\end{equation}
We note that the log-likelihood function associated with
the PDF $(\ref{eq:1pdf})$ can be classically written as
\begin{equation}
L(\bm{\Theta})=-\frac{N}{2}\Big(\ln\big[|\bm{R}_{\tilde{\bm{x}}}|\big]+\textrm{Tr}\big(\bm{R}_{\tilde{\bm{x}}}^{-1}\bm{R}_{\tilde{\bm{x}},N}\big)\Big)
\end{equation}
with $\bm{R}_{\tilde{\bm{x}},N}:=\frac{1}{N}\cdot\sum_{t=1}^{N}\tilde{\bm{x}}(t)\tilde{\bm{x}}^{H}(t)$, where $N$ is the snapshot number. Due to the structures of $\bm{R}_{\tilde{\bm{s}}}$ and $\bm{R}_{\tilde{\bm{n}}}$ in $\bm{R}_{\tilde{\bm{x}}}$, the ML estimation of $\bm{\Theta}$ can be obtained in a separable form. The ML estimation of $\bm{\vartheta}$ is got by minimizing with respect to parameter $\bm{\vartheta}$.
\begin{equation}
F_{N}(\bm{\vartheta})=\ln\big[\big|\tilde{\bm{A}}\hat{\bm{R}}_{\tilde{\bm{s}},ML}\tilde{\bm{A}}^{H}+\hat{\sigma}_{n,ML}^{2}\bm{I}_{4M}\big|\big]
\end{equation}
where $\hat{\bm{R}}_{\tilde{\bm{s}},ML}$, and $\hat{\sigma}_{n,ML}^{2}$ are given by
\begin{align}
\hat{\bm{R}}_{\tilde{\bm{s}},ML}=&\big[\tilde{\bm{A}}^{H}(\bm{\vartheta})\tilde{\bm{A}}(\bm{\vartheta})\big]^{-1}\tilde{\bm{A}}^{H}(\bm{\vartheta})\nonumber\\
&\cdot\big[\bm{R}_{\tilde{\bm{x}},N}-\hat{\sigma}_{n,ML}^{2}\bm{I}_{4M}\big]\tilde{\bm{A}}(\bm{\vartheta})\big[\tilde{\bm{A}}^{H}(\bm{\vartheta})\tilde{\bm{A}}(\bm{\vartheta})\big]^{-1}
\end{align}
and
\begin{equation}
\hat{\sigma}_{n,ML}^{2}=\frac{1}{2M-L}\textrm{Tr}\big(\bm{\Pi}_{\overline{\bm{A}}(\bm{\vartheta})}^{\perp}\bm{R}_{\bm{x},N}\big)
\end{equation}
where $\bm{\Pi}_{\overline{\bm{A}}(\bm{\vartheta})}$ is the projection matrix $\overline{\bm{A}}(\bm{\vartheta})\big[\overline{\bm{A}}^{H}(\bm{\vartheta})\overline{\bm{A}}(\bm{\vartheta})\big]^{-1}\overline{\bm{A}}^{H}(\bm{\vartheta})$.

Following the line of derivation given in \cite{Stoica1990Performance,delmas2004stochastic}, the CRB of parameter $\bm{\vartheta}$  is given by
\begin{align}
\bm{C}_{\bm{\vartheta}}=&\big[F^{\prime\prime}(\bm{\vartheta})\big]^{-1}\nonumber\\
&\cdot\Big(\lim_{N\to\infty}{\rm{E}}\big\{\big[F_{N}^{\prime}(\bm{\vartheta})\big]\big[F_{N}^{\prime}(\bm{\vartheta})\big]^{T}\big\}\Big)\big[F^{\prime\prime}(\bm{\vartheta})\big]^{-1}
\end{align}
where $F^{\prime}(\bm{\vartheta})$ is the gradient of $F_{N}(\bm{\vartheta})$, and $F^{\prime\prime}(\bm{\vartheta})$ is the limit of the Hessian of $F_{N}(\bm{\vartheta})$ when $N\to\infty$. The derivation details are given in the supplemental materials, then the CRB of non-circular signals $\bm{C}_{\bm{\vartheta}}^{(NC)}$ is given by

\begin{equation}
\begin{aligned}
\bm{C}_{\bm{\vartheta}}^{(NC)}=&\frac{\sigma_{n}^{2}}{2}\Big\{\textrm{Re}\Big[\bm{D}^{H}\bm{\Pi}_{\overline{\bm{A}}}^{\perp}\bm{D}\odot\Big(\bm{J}_{4}\otimes
\Big(\big[\bm{R}_{\bm{s}}\overline{\bm{A}}^{H},\bm{R}_{\bm{s}}^{\prime}\overline{\bm{A}}^{T}\big]\\
&\bm{R}_{\tilde{\bm{x}}}^{-1}\Big[
\begin{array}{c} \overline{\bm{A}}\bm{R}_{\bm{s}}\\ \overline{\bm{A}}^{*}\bm{R}_{\bm{s}}^{\prime*}\\ \end{array}
\Big]\Big)\Big)^{T}\Big]\Big\}^{-1}\in \mathbb{R}^{4L\times 4L},
\end{aligned}
\end{equation}
where
\begin{equation}
%\bm{\Pi}_{\bm{A}}^{\perp}:=&\bm{I}_{M}-\bm{A}[\bm{A}^{H}\bm{A}]^{-1}\bm{A}^{H}\\
\bm{D}:=\Big[\frac{\partial\overline{\bm{A}}}{\partial \bm{\vartheta}}\Big]\in \mathbb{C}^{2M\times 4L}
%=\Big[\frac{\partial\bm{\mathcal{a}}_{\ell}}{\partial \theta_{\ell}},\frac{\partial\bm{\mathcal{a}}_{\ell}}{\partial \psi_{\ell}}, \frac{\partial\bm{\mathcal{a}}_{\ell}}{\partial \gamma_{\ell}}, \frac{\partial\bm{\mathcal{a}}_{\ell}}{\partial \eta_{\ell}}\Big]\in \mathbb{C}^{M\times 4L}
\end{equation}

\vspace{-1em}
\section{Simulation Results}
In this section, we show numerical results to demonstrate the performance of the proposed algorithms. We compare the proposed QNC-MUSIC, DR-MUSIC algorithms with the ESPRIT (Only  $\theta$ and $\varphi$ is estimated by using the 2-D ESPRIT, the polarization parameter is based on spectral search which is similar as the QNC algorithm.) and Q-MUSIC  \cite{Abeida2006MUSIC}, the QDR-MUSIC algorithms introduced in section IV. A \cite{Li2011The}, respectively. The LV-MUSIC is not considered in our simulations because of its prohibitive computational complexity. For all simulations, we consider a URA equipped with EMVSs shown in Fig. \ref{Fig1}.  The horizontal inter-EMVS spacing $d_1$ and vertical inter-EMVS spacing $d_2$ are set to be $\lambda/2$. The data symbols are BPSK signal (non-circular signal) with unit power. The signal-to-noise ratio (SNR) is defined as $10\mathrm{log}_{10}(1/\sigma^2)$ where $\sigma^2$ is the noise power. The search step size  $0.1^\circ$ has been used for $\theta$ and $\varphi$; $0.03^\circ$ for $\gamma$ and $\eta$. The number of independent trials is 200. The metric of root-mean-square error (RMSE) is evaluated for the estimations of various source parameters.

\begin{figure}
  \centering
  \includegraphics[width=3.2in,height=2.4in]{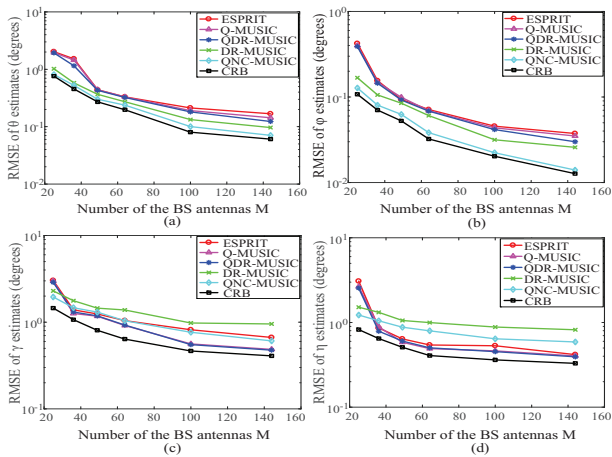}
\vspace{-1em}
\caption{RMSEs versus the number of the BS antennas $M$ for the estimates of DOA and polarization parameters when using different estimation algorithms, and the average received SNR from each mobile terminal is $0$ dB. (a), (b), (c) and (d) correspond to the estimation of azimuth, elevation, polarization angle and phase difference, respectively.}\label{Fig2}
 \vspace{-2em}
\end{figure}

\begin{figure}
  \centering
  \includegraphics[width=3.2in,height=2.4in]{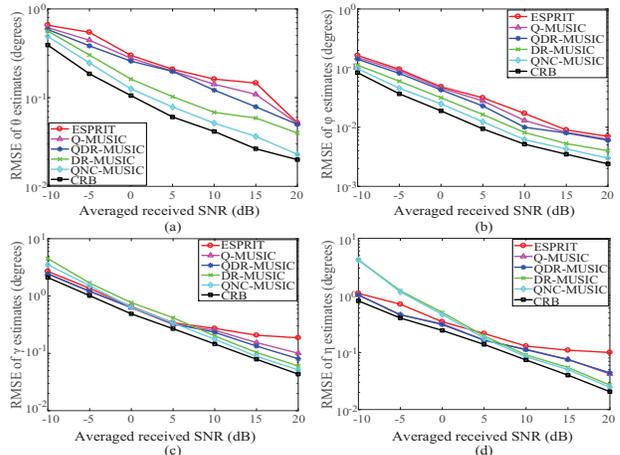}
\vspace{-1em}
\caption{RMSEs versus the average received SNR from each mobile terminal for the estimates of DOA and polarization parameters when using different estimation algorithms, while the number of the BS antennas is $M=64$, and the snapshot number is $N=500$. (a), (b), (c) and (d) correspond to the estimation of azimuth, elevation, polarization angle and phase difference, respectively.}\label{Fig3}
\vspace{-2em}
\end{figure}

The simulation parameters in the first two simulations as shown in Fig. \ref{Fig2} and Fig. \ref{Fig3} are given as follows. The number of the mobile terminals is $L=3$. The azimuth DOAs of three mobile terminals are $\overline{\theta}_1=20^\circ$, $\overline{\theta}_2=75^\circ$, $\overline{\theta}_3=115^\circ$, and the corresponding elevation DOAs are $\overline{\varphi}_1=10^\circ$, $\overline{\varphi}_2=15^\circ$, $\overline{\varphi}_3=20^\circ$; the corresponding polarization angles are $\overline{\gamma}_1=40^\circ$, $\overline{\gamma}_2=70^\circ$, $\overline{\gamma}_3=25^\circ$, and the corresponding phase differences are $\overline{\eta}_1=20^\circ$, $\overline{\eta}_2=40^\circ$, $\overline{\eta}_3=30^\circ$.

In the first test as shown in Fig. \ref{Fig2}, the average received SNR from each mobile terminal is $0$ dB, and the snapshot number is $N=200$. The number of EMVSs of the BS in the $x$-direction and the $y$-direction satisfy $M_x=M_y=\sqrt{M}$. The RMSEs of the estimated DOA and polarization parameter versus the number of the BS antennas $M$ are depicted in Fig. \ref{Fig2}. It can be observed that the RMSEs of these estimated parameters of the three algorithms decrease as $M$ increases. For the DOA estimation, the RMSEs of the DR-MUSIC are smaller than those of the Q-MUSIC and QDR-MUSIC. The reason is that the dimension of the covariance matrix $\hat{\bm{R}}_{LV}$ is larger than that of the covariance matrix $\hat{\bm{R}}$, the superiority of the LV-MUSIC has been reserved in the DR-MUSIC. The RMSEs of the QNC-MUSIC are smaller than those of the other algorithms and approach the CRB closely. This is because that the property of the non-circular signal has been used in the QNC-MUSIC to improve the accuracy of the DOA estimation. However, ESPRIT performs the worst of all, since the array aperture is not utilized completely. The RMSEs of the polarization parameter of the ESPRIT, the Q-MUSIC and the QDR-MUSIC are larger than those of the DR-MUSIC and the QNC-MUSIC with $M=25$, but the RMSEs of the polarization parameter of the ESPRIT, the Q-MUSIC and the QDR-MUSIC are smaller than those of the two algorithms as $M$ further increases. This is because that the polarization parameter estimation of the DR-MUSIC and the QNC-MUSIC has a closed-form expression, and the polarization parameter estimation of the ESPRIT, the Q-MUSIC and the QDR-MUSIC based on spectrum search exploiting the orthogonality between the signal and noise subspaces, i.e., $\|\bm{a}_{Ql}^H(\gamma ,\eta )\hat{\bm{U}}_{N}\|^2$. For the three algorithms, the estimation accuracy of the polarization parameter depends on not only the estimation accuracy of the DOA parameter, but also the dimension of the estimated noise subspace ${\hat{\bm{U}}}_N$. The dimension of ${\hat{\bm{U}}}_N$ increases as $M$ increases, and the orthogonality between the signal and noise subspace plays a more important role than the estimation accuracy of the DOA parameter in the polarization parameter estimation as $M$ increases. Thus the RMSEs of the polarization parameter are smaller than those of the other two algorithms as $M$ increases. However, as shown in the last section, the computational complexity of the QDR-MUSIC is about $LM^2J_3J_4-6M^2J_1J_2$ flops larger than that of the DR-MUSIC and the QNC-MUSIC. This condition would be even worse as the source number increases.

In the second test as shown in Fig. \ref{Fig3}, the number of EMVSs of the BS in the $x$-direction and the $y$-direction are $M_x=8$ and $M_y=8$, respectively, and hence $M=64$. The snapshot number is $N=200$. The RMSEs of the estimated DOA and polarization parameter versus the average received SNR from each mobile terminal are depicted in Fig. \ref{Fig3}. It can be observed that the RMSEs of these estimated parameters of the three algorithms decrease as SNR increases. For the DOA estimation, the RMSEs of the DR-MUSIC are smaller than those of the ESPRIT, the Q-MUSIC and the QDR-MUSIC, and QNC-MUSIC outperforms the other algorithms. The reason is identical with the first test. For the polarization parameter estimation, the RMSEs of the ESPRIT, the Q-MUSIC and the QDR-MUSIC are smaller than those of the DR-MUSIC and the QNC-MUSIC when the SNR is less than $0$ dB. This is caused by the fact that the orthogonality between the signal and noise subspaces plays a more important role than the estimation accuracy of the DOA parameter when the SNR is less than $0$ dB. The RMSEs of the ESPRIT, the Q-MUSIC and the QDR-MUSIC are larger than those of the DR-MUSIC and the QNC-MUSIC when the SNR is larger than $0$ dB. The reason is that the estimation accuracy of the DOA parameter is very high at a high SNR, and the effect of the orthogonality between the signal and noise subspaces is much weaker. These results demonstrate that the estimation accuracy of the polarization parameter deteriorates when the power of the received noise is high. It should be noted that the RMSEs of the polarization parameter become smaller as the number of the BS antennas increases. Therefore, the proposed DR-MUSIC and QNC-MUSIC can potentially achieve good performance by employing a larger number of EMVSs. In other words, for the polarized massive MIMO systems the transmitted power can be significantly reduced due to the application of an unprecedented large number of EMVS at the BS.

\begin{figure}
  \centering
  \includegraphics[width=3.2in,height=2.4in]{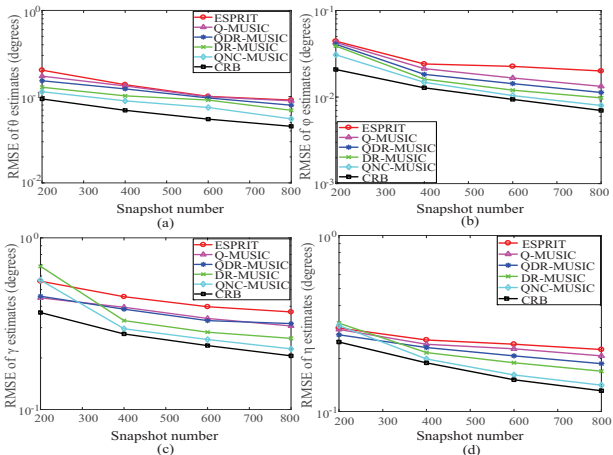}
  \vspace{-1em}
\caption{RMSEs versus the snapshot number $N$ received at the BS for the estimates of DOA and polarization parameters when using different estimation algorithms, while the number of the BS antennas is $M=64$, and the average received SNR from each mobile terminal is $5$ dB. (a), (b), (c) and (d) correspond to the estimation of azimuth, elevation, polarization angle and phase difference, respectively.}\label{Fig4}
\vspace{-2em}
\end{figure}

In the third test as shown in Fig. \ref{Fig4}, the number of EMVSs of the BS in the $x$-direction and the $y$-direction are $M_x=8$ and $M_y=8$, respectively, and hence $M=64$. The average received SNR from each mobile terminal is $5$ dB. The RMSEs of the estimated DOA and polarization parameter versus the snapshot number $N$ received at the BS are depicted in Fig. \ref{Fig4}. It can be observed that the RMSEs of these estimated parameters of the three algorithms decrease as the snapshot number increases. The curves shown in Fig. \ref{Fig4} are flatter than those shown in Fig. \ref{Fig3}, which means that for the parameter estimation performance, the effect of the snapshot number is slightly less than that of the SNR. For the polarization parameter estimation, the RMSEs of the ESPRIT, the Q-MUSIC and the QDR-MUSIC are smaller than those of the DR-MUSIC and the QNC-MUSIC when the snapshot number is less than 400; the RMSEs of the ESPRIT, the Q-MUSIC and the QDR-MUSIC are larger than those of the DR-MUSIC and the QNC-MUSIC when snapshot number is larger than 400. The effect of increasing the snapshot number is similar to the effect of increasing the SNR, as compared with Fig. \ref{Fig3}.

\begin{figure}
  \centering
  \includegraphics[width=3.2in,height=2.4in]{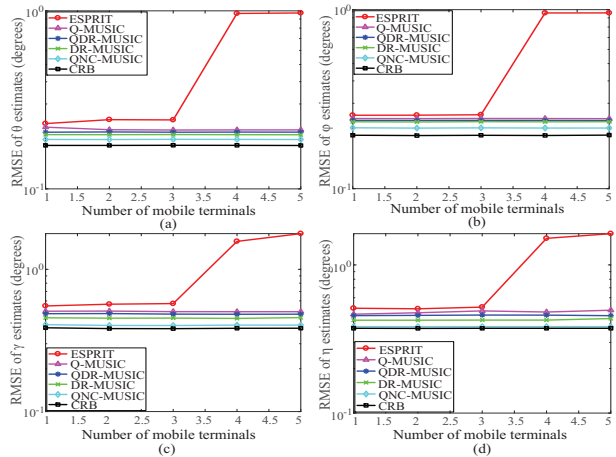}
\vspace{-1em}
\caption{RMSEs versus the number of the mobile terminals for the estimates of DOA and polarization parameters when using different estimation algorithms, while the number of the BS antennas is $M=100$. The average received SNR from each mobile terminal is $0$ dB, and the snapshot number is $N=500$. (a), (b), (c) and (d) correspond to the estimation of azimuth, elevation, polarization angle and phase difference, respectively.}\label{Fig5}
\vspace{-2em}
\end{figure}

In the fourth test as shown in Fig. \ref{Fig5}, some of the parameters are changed for evaluating the performance of the three algorithms with the increased number of mobile terminals. A more realistic scenario considering the multipath propagation in 3D millimeter wave channels is used for simulation. For each mobile terminal, the polarization parameter is not changed, but the DOA parameters are different. There are $3$ dominant paths from each mobile terminal to the BS, and the signal among them are coherent with each other. For the first mobile terminal, the polarization angle is $\overline{\gamma}_1=40^\circ$, and the phase difference is $\overline{\eta}_1=20^\circ$; the corresponding azimuth DOAs are $\overline{\theta}_{11}=20^\circ$, $\overline{\theta}_{12}=30^\circ$, $\overline{\theta}_{13}=50^\circ$, and the corresponding elevation DOAs are $\overline{\varphi}_{11}=10^\circ$, $\overline{\varphi}_{12}=15^\circ$, $\overline{\varphi}_{13}=20^\circ$. For the second mobile terminal, the polarization angle is $\overline{\gamma}_2=10^\circ$, and the phase difference is $\overline{\eta}_2=60^\circ$; the corresponding azimuth DOAs are $\overline{\theta}_{21}=25^\circ$, $\overline{\theta}_{22}=60^\circ$, $\overline{\theta}_{23}=125^\circ$, and the corresponding elevation DOAs are $\overline{\varphi}_{21}=15^\circ$, $\overline{\varphi}_{22}=50^\circ$, $\overline{\varphi}_{23}=40^\circ$. For the third mobile terminal, the polarization angle is $\overline{\gamma}_3=25^\circ$, and the phase difference is $\overline{\eta}_3=30^\circ$; the corresponding azimuth DOAs are $\overline{\theta}_{31}=40^\circ$, $\overline{\theta}_{32}=80^\circ$, $\overline{\theta}_{33}=110^\circ$, and the corresponding elevation DOAs are $\overline{\varphi}_{31}=25^\circ$, $\overline{\varphi}_{32}=60^\circ$, $\overline{\varphi}_{33}=35^\circ$. For the fourth mobile terminal, the polarization angle is $\overline{\gamma}_4=70^\circ$, and the phase difference is $\overline{\eta}_4=40^\circ$; the corresponding azimuth DOAs are $\overline{\theta}_{41}=70^\circ$, $\overline{\theta}_{42}=125^\circ$, $\overline{\theta}_{43}=140^\circ$, and the corresponding elevation DOAs are $\overline{\varphi}_{41}=30^\circ$, $\overline{\varphi}_{42}=45^\circ$, $\overline{\varphi}_{43}=60^\circ$. For the fifth mobile terminal, the polarization angle is $\overline{\gamma}_5=80^\circ$, and the phase difference is $\overline{\eta}_5=50^\circ$; the corresponding azimuth DOAs are $\overline{\theta}_{51}=75^\circ$, $\overline{\theta}_{52}=130^\circ$, $\overline{\theta}_{53}=150^\circ$, and the corresponding elevation DOAs are $\overline{\varphi}_{51}=20^\circ$, $\overline{\varphi}_{52}=55^\circ$, $\overline{\varphi}_{53}=75^\circ$.

The number of EMVSs of the BS in the $x$-direction and the $y$-direction are $M_x=10$ and $M_y=10$, respectively, and hence $M=100$. The average received SNR from each mobile terminal is $0$ dB, and the snapshot number is $N=500$. The 2-D spatial smoothing technique has been used for solving the coherent signals, and the size of subarray is $8\times8$. The RMSEs of the estimated DOA and polarization parameter versus the number of the mobile terminals are depicted in Fig. \ref{Fig5}. It can be observed that the RMSEs of the five algorithms except ESPRIT increase slowly as the number of the mobile terminal increases. This is because a rough estimation of DOA parameter and DOA, polarization parameter have been obtained for the DR-MUSIC, QNC-MUSIC and the Q-MUSIC, the QDR-MUSIC, respectively,  based on a large step size of the spectrum search. For the Q-MUSIC and the QDR-MUSIC, the DOA and polarization parameter of the mobile terminals are only estimated by searching around the true values. For the DR-MUSIC and the QNC-MUSIC, the DOA parameters of the mobile terminals are only estimated by searching around the true values, while the polarization parameter has a closed form. Thus the RMSEs of these four algorithms change slowly. For the ESPRIT, the RMSEs increases greatly compared with the other four algorithms. This is mainly because that the array aperture is not utilized completely. Some mobile terminals are too close in space, the ESPRIT cannot distinguish them exactly, thus the RMSEs increases as the number of mobile terminals increases.

\vspace{-1em}
\section{Conclusion}
In this paper, we have proposed a MUSIC-based algorithm, QNC-MUSIC, for the 2-D DOA and polarization estimation of the non-circular signal in the 3-D millimeter wave polarized massive MIMO systems. For the DOA estimation of non-circular signals using the QNC-MUSIC, the property of non-circular signals is used to further improve the DOA estimation accuracy. For the polarization estimation of the non-circular signal, the closed-form expression of the DR-MUSIC is adopted based on the DOA estimation result of the QNC-MUSIC. Compared with the traditional LV-MUSIC, Q-MUSIC and the QDR-MUSIC, the computational complexity of the QNC-MUSIC and the DR-MUSIC are much lower with the help of the derivative of the spectrum function and the closed-form expression of the polarization estimation. Our analysis and simulation show that the performance of the proposed QNC-MUSIC improves as the number of the BS antennas increases, and the DOA estimation accuracy is higher than other algorithms in massive MIMO systems in particular. In the future, the method of utilizing the property of non-circular signals to improve the estimation performance of the polarization parameter should be studied.

%\vspace{-1em}

%\section*{Acknowledgment}

%\begin{IEEEbiography}[{\includegraphics[width=1in,height=1.25in,clip,keepaspectratio]{LiangtianWan.eps}}]{Liangtian Wan}(M'15)
%received the B.S. degree and the Ph.D. degree in the College of Information and Communication Engineering from Harbin Engineering University, Harbin, China, in 2011 and 2015, respectively. From Oct. 2015 to Apr. 2017, he has been a Research Fellow of School of Electrical and Electrical Engineering, Nanyang Technological University, Singapore. He is currently an Associate Professor of School of Software, Dalian University of Technology, China. He is the author of over 30 papers published in related international conference proceedings and journals. Dr. Wan has been serving as an Associate Editor (a Guest Editor) for IEEE Access and Journal of Information Processing Systems. His current research interests include social network analysis and mining, big data, array signal processing, wireless sensor networks, compressive sensing.
%\end{IEEEbiography}

%\begin{IEEEbiography}[{\includegraphics[width=1in,height=1.25in,clip,keepaspectratio]{TongZhu.eps}}]{Tong Zhu}
%received the B.S. degree and the Ph.D. degree in the College of Information and Communication Engineering from Harbin Engineering University, Harbin, China, in 2010 and 2014, respectively. He is currently a Researcher at Tianjin Institute of Computing Technology, China. His research interests include array signal processing and statistical signal pcessing.
%\end{IEEEbiography}

\vspace{-1em}
\begin{thebibliography}{1}
\bibitem{Akdeniz2014Millimeter}
M. R. Akdeniz,  Y. Liu, M. K. Samimi, S. Sun, S. Rangan, T. S. Rappaport, and Elza Erkip, `` Millimeter wave channel modeling and cellular capacity evaluation,"\hskip 1em plus 0.5em minus 0.4em\relax  \emph{IEEE J. Sel. Areas Commun.}, vol. 32, no. 6, pp. 1164--1179, Jun. 2014.
\bibitem{Marzi2016Compressive}
Z. Marzi, D. Ramasamy, and U. Madhow, ``Compressive channel estimation and tracking for large arrays in mm-Wave picocells,"\hskip 1em plus 0.5em minus 0.4em\relax  \emph{IEEE J. Sel. Topics Signal Process.}, vol. 10, no. 3, pp. 514--527, Apr. 2016.
\bibitem{Shafin2016DoA}
R. Shafin, L. Liu, C. Zhang, and Y. C. Wu, ``DoA estimation and capacity analysis for 3D millimeter wave massive-MIMO/FD-MIMO OFDM systems,"\hskip 1em plus 0.5em minus 0.4em\relax  \emph{IEEE Trans. Wireless Commun.},  vol. 15, no. 10, pp. 6963--6978, Oct. 2016.
\bibitem{Larsson2014Massive}
E. G. Larsson, O. Edfors, F. Tufvesson, and T. L. Marzetta, ``Massive MIMO for next generation wireless systems,"\hskip 1em plus 0.5em minus 0.4em\relax  \emph{IEEE Commun. Mag.},  vol. 52, no. 2, pp. 186--195, Feb. 2014.
\bibitem{Cheng2015Subspace}
L. Cheng, Y. C. Wu, J. Zhang, and L. Liu, ``Subspace identification for DOA estimation in massive/full-dimension MIMO systems: bad data mitigation and automatic source enumeration,"\hskip 1em plus 0.5em minus 0.4em\relax  \emph{IEEE Trans. Signal Process.},  vol. 63, no. 22, pp. 5897--5909, Nov. 15, 2015.
\bibitem{Yin2013A}
H. Yin, D. Gesbert, M. Filippou, and Y. Liu, ``A coordinated approach to channel estimation in large-scale multiple-antenna systems,"\hskip 1em plus 0.5em minus 0.4em\relax  \emph{IEEE J. Sel. Areas Commun.},  vol. 31, no. 2, pp. 264--273, Feb. 2013.
\bibitem{He2014Leakage}
S. He,  Y. Huang,  H. Wang,  S. Jin, and  L. Yang, ``Leakage-aware energy-efficient beamforming for heterogeneous multicell multiuser systems,"\hskip 1em plus 0.5em minus 0.4em\relax  \emph{IEEE J. Sel. Areas Commun.},  vol. 32, no. 6, pp. 1268--1281, Jun. 2014.
\bibitem{Adhikary2013Joint}
A. Adhikary, J. Nam, J.-Y. Ahn, and G. Caire, ``Joint spatial division and multiplexing: The large-scale array regime,"\hskip 1em plus 0.5em minus 0.4em\relax  \emph{IEEE Trans. Inf. Theory},  vol. 59, no. 10, pp. 6441--6463, Oct. 2013.
\bibitem{Alkhateeb2014Channel}
A. Alkhateeb, O. El Ayach, G. Leus, and R. W. Heath, Jr., ``Channel estimation and hybrid precoding for millimeter wave cellular systems,"\hskip 1em plus 0.5em minus 0.4em\relax  \emph{IEEE J. Sel. Topics Signal. Process.}, vol. 8, no. 5, pp. 831--846, Oct. 2014.
\bibitem{heath2016overview}
R. W. Heath, Jr., N. Gonz\'{a}lez-Prelcic, S. Rangan, W. Roh, and A. M. Sayeed, ``An overview of signal processing techniques for millimeter wave MIMO systems,"\hskip 1em plus 0.5em minus 0.4em\relax  \emph{IEEE J. Sel. Topics Signal Process.}, vol. 10, no. 3, pp. 436--453, Oct. 2016.
\bibitem{Andrews2017Modeling}
J. G. Andrews, T. Bai, M. N. Kulkarni, A. Alkhateeb, A. K. Gupta,
and R. W. Heath, Jr., ``Modeling and analyzing millimeter wave cellular
systems,"\hskip 1em plus 0.5em minus 0.4em\relax  \emph{IEEE Trans. Commun.}, vol. 65, no. 1, pp. 403¨C-430, Jan. 2017.
%\bibitem{Zhou2016Channel}
%Z. Zhou, J. Fang, L. Yang, H. Li, Z. Chen, and S. Li, ``Channel Estimation for Millimeter Wave Multiuser MIMO Systems via PARAFAC Decomposition,"\hskip 1em plus 0.5em minus 0.4em\relax  \emph{IEEE Trans. Wireless Commun.},  vol. 15, no. 11, pp. 7501--7516, Nov. 2016.
%\bibitem{Zhou2017Low}
%Z. Zhou, J. Fang, L. Yang, H. Li, Z. Chen, and R. S. Blum, ``Low-Rank tensor decomposition-aided channel estimation for millimeter wave MIMO-OFDM systems,"\hskip 1em plus 0.5em minus 0.4em\relax  \emph{IEEE J. Sel. Areas Commun.},  vol. 35, no. 7, pp. 1524--1538, Jul. 2017.
\bibitem{Gounon1998Localisation}
P. Gounon,  C. Adnet and J. Galy , ``Localisation angulaire de signaux non circulaires,"\hskip 1em plus 0.5em minus 0.4em\relax  \emph{Trait. Signal}, vol. 15, no. 1, pp. 17--23, 1998.
\bibitem{Abeida2006MUSIC}
H. Abeida and J. P. Delmas, ``MUSIC-like estimation of direction of arrival for noncircular sources,"\hskip 1em plus 0.5em minus 0.4em\relax  \emph{IEEE Trans. Signal Process.}, vol. 54, no. 7, pp. 2678--2690 , Jul. 2006.
\bibitem{Delmas2004Asymptotically}
J. P. Delmas, ``Asymptotically minimum variance second-order estimation for noncircular signals with application to DOA estimation,"\hskip 1em plus 0.5em minus 0.4em\relax  \emph{IEEE Trans. Signal Process.}, vol. 52, no. 5, pp. 1235--1241, May 2004.
\bibitem{Charge2001A}
P. Charge, Y. Wang and J. Saillard, ``A root-MUSIC algorithm for non circular sources,"\hskip 1em plus 0.5em minus 0.4em\relax in \emph{Proc. IEEE Int. Conference on Acoustics, Speech, and Signal Processing (ICASSP)}, Salt Lake City, USA, May. 2001.
\bibitem{Liu2012A}
A. Liu, G. Liao, Q. Xu and C. Zeng, ``A circularity-based DOA estimation method under coexistence of noncircular and circular signals,"\hskip 1em plus 0.5em minus 0.4em\relax in \emph{Proc. IEEE Int. Conference on Acoustics, Speech, and Signal Processing (ICASSP)},Kyoto, Japan, Mar. 2012.
\bibitem{Gao2013Improved}
F. Gao, A. Nallanathan and Y. Wang, ``Improved MUSIC under the coexistence of both circular and noncircular sources,"\hskip 1em plus
  0.5em minus 0.4em\relax  \emph{IEEE Trans. Signal Process.}, vol. 56, no. 7, pp. 3033--3038, Jul. 2008.
\bibitem{Gou2013Biquaternion}
X. Gou, Z. Liu, and Y. Xu, ``Biquaternion cumulant-MUSIC for DOA estimation of noncircular signals,"\hskip 1em plus 0.5em minus 0.4em\relax  \emph{Signal Process.},  vol. 93, no. 4, pp. 874--881, Apr. 2013.
%\bibitem{IEEEhowto:6}
%H. Abeida and J. P. Delmas, ``MUSIC-like estimation of direction of arrival for noncircular sources,"\hskip 1em plus 0.5em minus 0.4em\relax  \emph{IEEE Trans. Signal Process.}, vol. 54, no. 7, pp. 2678--2690 , Jul. 2006.
%\bibitem{IEEEhowto:7}
%L. Wan, G. Han, J. Jiang, J. J. P. C. Rodrigues, N. Feng and T. Zhu, ``DOA estimation for coherently distributed sources considering circular and noncircular signals in massive MIMO systems,"\hskip 1em plus 0.5em minus 0.4em\relax  \emph{IEEE Syst. J.}, 2015, to be published. DOI: 10.1109/JSYST.2015.2445052.
%\bibitem{IEEEhowto:8}
%H. Abeida and J. P. Delmas, ``Statistical performance of MUSIC-Like algorithms in resolving noncircular sources,"\hskip 1em plus 0.5em minus 0.4em\relax  \emph{IEEE Trans. Signal Process.}, vol. 56, no. 9, pp. 4317--4329 , Sep. 2008.
%\bibitem{IEEEhowto:9}
%J. P. Delmas and H. Abeida, ``Stochastic Cram\'{e}r-Rao bound for noncircular signals with application to DOA estimation,"\hskip 1em plus 0.5em minus 0.4em\relax  \emph{IEEE Trans. Signal Process.}, vol. 56, no. 9, pp. 4317--4329 , Sep. 2008.
\bibitem{Liu2012Direction}
Z. M. Liu, Z. T. Huang, Y. Y. Zhou, and J. Liu, ``Direction-of-arrival estimation of noncircular signals via sparse representation,"\hskip 1em plus 0.5em minus 0.4em\relax  \emph{IEEE Trans. Aerosp. Electron. Syst.}, vol. 48, no. 3, pp. 2690--2698, Jul. 2012.
\bibitem{Zoubir2003Non}
A. Zoubir, P. Charg\'{e}, and Y. Wang, ``Non circular sources localization with ESPRIT,"\hskip 1em plus 0.5em minus 0.4em\relax  in \emph{Eur. Conf. Wireless Technol. (ECWT)}, Munich, Germany, Oct. 2003.
\bibitem{Haardt2004Enhancements}
M. Haardt and F. Roemer, ``Enhancements of unitary ESPRIT for noncircular sources,"\hskip 1em plus 0.5em minus 0.4em\relax  in \emph{IEEE Int. Conf. Acoust., Speech,
Signal Processing (ICASSP)}, Montreal, QC, Canada, May 2004.
\bibitem{Roemer2014Analytical}
F. Roemer,  M. Haardt, and  G. Del Galdo, ``Analytical performance assessment of multi-dimensional matrix- and tensor-based ESPRIT-type algorithms,"\hskip 1em plus 0.5em minus 0.4em\relax  \emph{IEEE Trans. Signal Process.}, vol. 62, no. 10, pp. 2611--2625 , May 2014.
\bibitem{Steinwandt2014R}
J. Steinwandt, F. Roemer, M. Haardt, and G. Del Galdo, ``R dimensional ESPRIT-type algorithms for strictly second-order noncircular sources and their performance analysis,"\hskip 1em plus 0.5em minus 0.4em\relax  \emph{IEEE Trans. Signal Process.}, vol. 62, no. 18, pp. 4824--4838, Sep. 2014.
\bibitem{Steinwandt2015ESPRIT}
J. Steinwandt, F. Roemer, and M. Haardt, ``ESPRIT-Type Algorithms for a Received Mixture of Circular and Strictly Non-Circular Signals,"\hskip 1em plus 0.5em minus 0.4em\relax in \emph{Proc. IEEE Int. Conf. Acoustics, Speech and Sig. Proc. (ICASSP 2015)}, Brisbane, Australia, Apr. 2015.
\bibitem{Roemer2006Efficient}
F. Roemer and M. Haardt, ``Efficient 1-D and 2-D DOA estimation for non-circular sources with hexagonal shaped espar arrays,"\hskip 1em plus
  0.5em minus 0.4em\relax  in \emph{Proc. IEEE Int. Conference on Acoustics, Speech, and Signal Processing (ICASSP)}, Toulouse, France, pp. 881--884, May 2006.
\bibitem{Li1991Angle}
J. Li and R. T. C. Jr, ``Angle and polarization estimation using ESPRIT with a polarization sensitive array,"\hskip 1em plus
  0.5em minus 0.4em\relax  \emph{IEEE Trans. Antennas Propag.}, vol. 39, no. 9, pp. 1376--1383, Sep. 1991.
\bibitem{Wong1996Diversely}
K. T. Wong and M. D. Zoltowski, ``Diversely polarized root-MUSIC for azimuth- elevation angle of arrival estimation,"\hskip 1em plus
  0.5em minus 0.4em\relax  \emph{Dig. 1996 IEEE Antennas Propagation Soc. Int. Symp.}, pp. 1352--1355, Sep. 1996.
\bibitem{Gonen1999Applications}
E. Gonen and J. M. Mendel, ``Applications of cumulants to array processing. Part VI. Polarization and direction of arrival estimation with minimally constrained arrays,"\hskip 1em plus 0.5em minus 0.4em\relax  \emph{IEEE Trans. Signal Process.}, vol. 47, no. 9, pp. 2589--2592, Sep. 1999.
\bibitem{Chevalier2007Higher}
P. Chevalier, A. Ferr\'{e}ol, L. Albera, and Gw\'{e}na\"{e}l Birot, ``Higher order direction finding from arrays with diversely polarized antennas: the PD-2q-MUSIC algorithms,"\hskip 1em plus 0.5em minus 0.4em\relax  \emph{IEEE Trans. Signal Process.}, vol. 55, no. 11, pp. 5337--5350, Nov. 2007.
\bibitem{Liu2001Cramer}
X. Liu and N. D. Sidiropoulos, ``Cram\'{e}r-Rao lower bounds for low-rank decomposition of multidimensional arrays,"\hskip 1em plus 0.5em minus 0.4em\relax  \emph{IEEE Trans. Signal Process.}, vol. 49, no. 9, pp. 2074--2086, Sep. 2001.
\bibitem{Guo2011A}
X. Guo, S. Miron, D. Brie, S. Zhu, and X. Liao, ``A CANDECOMP/PARAFAC perspective on uniqueness of DOA estimation using a vector sensor array,"\hskip 1em plus 0.5em minus 0.4em\relax  \emph{IEEE Trans. Signal Process.}, vol. 59, no. 7, pp. 3475--3481, Jul. 2011.
\bibitem{Costa2012DoA}
M. Costa, A. Richter, and V. Koivunen, ``DoA and polarization estimation for arbitrary array configurations,"\hskip 1em plus 0.5em minus 0.4em\relax  \emph{IEEE Trans. Signal Process.}, vol. 60, no. 5, pp. 2330--2343, May 2012.
\bibitem{Tian2015Sparse}
Y. Tian, X. Sun, S. Zhao, ``Sparse-reconstruction-based direction of arrival, polarisation and power estimation using a cross-dipole array,"\hskip 1em plus 0.5em minus 0.4em\relax  \emph{IET Radar Sonar Navig.}, vol. 9, no. 6, pp. 727--731, Jul. 2015.
\bibitem{Miron2006Quaternion}
S. Miron, N. L. Bihan, and J. I. Mars, ``Quaternion-MUSIC for vector-sensor array processing,"\hskip 1em plus 0.5em minus 0.4em\relax  \emph{IEEE Trans. Signal Process.}, vol. 54, no. 4, pp. 1218--1229, Apr. 2006.
\bibitem{Bihan2013MUSIC}
N. L. Bihan, S. Miron, and J. I. Mars, ``MUSIC algorithm for vector-sensors array using biquaternions,"\hskip 1em plus 0.5em minus 0.4em\relax  \emph{IEEE Trans. Signal Process.}, vol. 55, no. 9, pp. 4523--4533, Sep. 2013.
\bibitem{Gong2011Direction}
X. Gong, Z. W. Liu, and Y. G. Xu, ``Direction finding via biquaternion matrix diagonalization with vector-sensors,"\hskip 1em plus 0.5em minus 0.4em\relax  \emph{Signal Process.}, vol. 91, no. 4, pp. 821--831, Apr. 2011.
\bibitem{Poon2011Degree}
A. S. Y. Poon and D. N. C. Tse, ``Degree-of-freedom gain from using polarimetric antenna elements,"\hskip 1em plus 0.5em minus 0.4em\relax  \emph{IEEE Trans. Inf.
Theory},  vol. 57, no. 9, pp. 5695--5709, Sep. 2011.
\bibitem{Su2016Channel}
X. Su, D. Choi, X. Liu, and B Peng, ``Channel Model for Polarized MIMO Systems With Power Radiation Pattern Concern,"\hskip 1em plus 0.5em minus 0.4em\relax  \emph{IEEE Access},  vol. 4, pp. 1061--1072, Mar. 2016.
\bibitem{Li2011The}
J. Li and J. Tao, ``The dimension reduction quaternion MUSIC algorithm for jointly estimating DOA and polarization,"\hskip 1em plus 0.5em minus 0.4em\relax  \emph{J. Electron. Inf. Technol.}, vol. 33, no. 1, pp. 106--111, Jan. 2011.
\bibitem{IEEEhowto:31}
W. Si, T. Zhu, and M. Zhang, ``Dimension-reduction MUSIC for jointly estimating DOA and polarization using plane polarized arrays,"\hskip 1em plus 0.5em minus 0.4em\relax  \emph{J. Commun.}, vol. 35, no. 12, pp. 28--35, Dec. 2014.
\bibitem{Bihan2004Singular}
N. L. Bihan and J. Mars, ``Singular value decomposition of quaternion matrices: a new tool for vector-sensor signal processing,"\hskip 1em plus 0.5em minus 0.4em\relax  \emph{ Signal Process.}, vol. 84, no. 7, pp. 1177--1199, Jul. 2004.
\bibitem{Ward1997Quaternions}
J. P. Ward, \hskip 1em plus 0.5em minus 0.4em\relax  \emph{Quaternions and Cayley Numbers, Algebra and Applications}. Norwell, MA: Kluwer, 1997.
\bibitem{Zhang1997Quaternions}
F. Zhang, ``Quaternions and matrices of quaternions," \hskip 1em plus 0.5em minus 0.4em\relax  \emph{Linear Algebra Its Appl.}, vol. 251, pp. 21--57, Jan. 1997.
\bibitem{Schmidt1986Multiple}
R. Schmidt, ``Multiple emitter location and signal parameter estimation," \hskip 1em plus 0.5em minus 0.4em\relax  \emph{IEEE Trans. Antennas Propag.}, vol. 34, no. 3, pp. 276--280, Mar. 1986.
\bibitem{Xu1994Fast}
G. Xu and  T. Kailath, ``Fast subspace decomposition,"\hskip 1em plus 0.5em minus 0.4em\relax  \emph{IEEE Trans. Signal Process.}, vol. 42, no. 3, pp. 539--551, Mar. 1994.
\bibitem{Stoica1990Performance}
P. Stoica and A. Nehorai, ``Performance study of conditional and unconditional direction-of-arrival estimation,"\hskip 1em plus 0.5em minus 0.4em\relax  \emph{IEEE Trans. Acoust., Speech, Signal Process.}, vol. 38, no. 10, pp. 1783--1795, Oct. 1990.
\bibitem{Stoica2001The}
P. Stoica, E. G. Larsson and A. B. Gershman, ``The stochastic CRB for array processing: a textbook derivation,"\hskip 1em plus 0.5em minus 0.4em\relax  \emph{IEEE Signal Process. Lett.}, vol. 8, no. 5, pp. 148--150, May 2001.
\bibitem{delmas2004stochastic}
H. Abeida and J. P. Delmas, ``Stochastic Cram\'{e}r-Rao Bound for noncircular signals with application to DOA estimation,"\hskip 1em plus 0.5em minus 0.4em\relax  \emph{IEEE Trans. Signal Process.}, vol. 52, no. 11, pp. 3192--3199, Nov. 2004.
\bibitem{hassen2011doa}
S. B. Hassen, F. Bellili, A. Samet and S. Affes, ``DOA estimation of temporally and spatially correlated narrowband noncircular sources in spatially correlated white noise,"\hskip 1em plus 0.5em minus 0.4em\relax  \emph{IEEE Trans. Signal Process.}, vol. 59, no. 9, pp. 4108--4121, Sep. 2011.
\end{thebibliography}
\end{document}